\documentclass[
superscriptaddress,
twocolumn,
amsmath,amssymb,
aps,
prb,
floatfix,
showkeys,
reprint,
]{revtex4-2}

\usepackage{graphicx}
\usepackage{dcolumn}
\usepackage{bm}
\usepackage{amsmath}
\usepackage{amssymb}
\usepackage{multirow}
\usepackage{xcolor}
\usepackage{booktabs}
\usepackage{tabularx}
\usepackage[english]{babel}
\usepackage[utf8]{inputenc}
\usepackage{enumerate}
\usepackage{newunicodechar}
\newunicodechar{ﬁ}{fi}
\newunicodechar{ﬀ}{ff}
\usepackage[colorlinks, citecolor=blue, linkcolor=blue]{hyperref}

\newcommand{\KIEL}{Institute of Theoretical Physics and Astrophysics, University of Kiel, Leibnizstrasse 15, 24098 Kiel, Germany}

\begin{document}
\title{Trends of higher-order exchange interactions in transition-metal trilayers }

\author{Mara Gutzeit}
\affiliation{\KIEL}

\author{Soumyajyoti Haldar}
\affiliation{\KIEL}

\author{Sebastian Meyer}
\altaffiliation[Current address: ]{Nanomat/Q-mat/CESAM, Universit{\'e} de Li{\`e}ge, B-4000 Sart Tilman, Belgium}
\affiliation{\KIEL}

\author{Stefan Heinze}
\affiliation{\KIEL}

\date{\today}

\begin{abstract}
We present a systematic study of higher-order exchange interactions beyond the pair-wise Heisenberg exchange in transition-metal trilayers based on density functional theory calculations. We show that these terms can play an important role in magnetic trilayers composed of a single hexagonal Fe or Co atomic layer sandwiched between $4d$ and $5d$ transition-metal layers. We study the dependence of the biquadratic and the three-site and four-site four spin interaction on the band filling
of the $4d$ and $5d$ layers as well as the stacking sequence, i.e.~fcc vs.~hcp stacking. Our calculations reveal relatively small higher-order interactions for Co based trilayers. For Fe based trilayers with a Rh or Ir layer the higher-order 
terms can be on the same order of magnitude as pair-wise Heisenberg exchange. The trends obtained for freestanding trilayers are used to understand the higher-order interactions in ultrathin film systems on surfaces that 
are experimentally accessible. 
It is shown that hcp-Rh/Fe/Ir(111) and hcp-Rh/Fe/Rh(111) exhibit the largest values for the biquadratic and the 
three-site four spin interaction of all systems under study. 
We further demonstrate that the three-site four spin interaction is responsible for the experimentally 
observed change of the magnetic ground state of Rh/Fe/Ir(111) from a spin spiral (single-Q) for fcc-Rh to a 2Q state for 
hcp-Rh. We find similar trends for Rh/Fe/Rh(111), i.e.~replacing the Ir surface
by the isoelectronic Rh surface. For Rh/Co/Ir(111), we obtain a negative value for the four-site four spin interaction which will lead to a reduced stability of magnetic skyrmions which are metastable in this film at zero magnetic field.
In contrast, for Pd/Fe/Ir(111), the four-site four spin interaction is positive which leads to an enhanced stability
of skyrmions.
\end{abstract}

\maketitle

\section{Introduction}
The magnetic properties of a material are commonly discussed starting from the Heisenberg model which rests on the
exchange interaction between pairs of magnetic moments. For metals, the pair-wise exchange
interaction can be obtained in second order perturbation theory from the Hubbard model which describes electrons on 
a lattice by their hopping between sites and their mutual Coulomb repulsion. However, spin interactions beyond the
conventional Heisenberg exchange arise if one includes higher-order terms in the perturbative expansion. It has been
shown that the two-site four spin (biquadratic) and the four-site four spin interaction occur in fourth order for
the spin-$1/2$ Hubbard model \cite{Takahashi_1977,MacDonald88}. Recently, it has been demonstrated that in addition a 
three-site four spin interaction arises if one considers systems with a spin $S\ge 1$ \cite{Hoffmann2020} appropriate 
in spin models for $3d$ transition-metals such as Co, Fe, or Mn with magnetic moments on the order of 2 to 3~$\mu_B$.
  
Higher-order exchange interactions (HOI) beyond the conventional pair-wise Heisenberg exchange can play a crucial role 
for the magnetic properties of transition-metal nanostructures. A model class of material systems consists of 
monolayers of magnetic $3d$ transition metals on $4d$ and $5d$ transition metal surfaces allowing a 
pseudomorphic growth of the $3d$ monolayer and experimental accessibility~\cite{Albrecht2000,Repetto2006}.
Prominent examples of complex noncollinear magnetic ground states driven by HOI
are the triple-Q state first proposed by Kurz {\it et al.} \cite{Kurz2001} and recently experimentally
observed in a Mn monolayer on the Re(0001) surface \cite{Spethmann2020} as well as the conical spin spiral ground state
of a Mn double layer on W(110) \cite{Yoshida2012}. The interplay with the Dzyaloshinskii-Moriya interaction (DMI), which 
occurs due to spin-orbit coupling 
in systems with broken inversion symmetry, 
even leads to the formation of a nanoskyrmion lattice in an Fe monolayer
on Ir(111) \cite{Heinze2011}. 
The coupling of spin spirals via HOI can also result in collinear magnetic ground states such as the up-up-down-down 
($uudd$) or double-row wise antiferromagnetic state \cite{Hardrat2009,Al-Zubi2011}. Recently, it has been shown that
the $uudd$ state can be triggered by the three-site four spin interaction \cite{Hoffmann2020}
and it was
experimentally observed in an Fe monolayer on the Rh(111) surface \cite{Kroenlein2018}. 

The magnetic properties in such systems can be tuned by adding another interface, e.g.~by growing a non-magnetic overlayer
on the $3d$ transition-metal films as in the Rh/Fe atomic bilayer on Ir(111) \cite{Romming2018}. In the latter system,
there is a surprising dependence of the magnetic ground state on the stacking of the Rh overlayer. While a spin spiral
state has been found for fcc-Rh stacking, a canted $uudd$ state exists for hcp-Rh stacking \cite{Romming2018}. The
canting is induced by the competition of the HOI and the DMI, however, the origin of the stacking dependence in terms
of the involved HOI has not yet been explained. Recently, a stacking-dependent change of the sign of the four-site four 
spin interaction has been reported for a Pd/Fe atomic bilayer on the Re(0001) surface \cite{Li2020}. The four-site four
spin interaction plays an important role for the stability of topological spin structures such as skyrmions and antiskyrmions 
which
has been exemplified for Pd/Fe bilayers on Rh(111) and Ir(111) \cite{Paul2020}. 

So far a systematic study about HOI at transition-metal interfaces is missing and only the Fe and Mn based ultrathin film systems mentioned above have been a matter of both theoretical and experimental research. 
Although isolated nanometer skyrmions were observed in Rh/Co bilayers on Ir(111) as well~\cite{Meyer2019}, no studies for HOI terms which could potentially contribute to the stabilization of these particle-like spin structures in Co based systems are available yet.
However, a more systematic investigation would shed light on the question whether these systems can be regarded as special or if there exits a trend indicating which types of transition-metal surfaces might show large effects on the HOI terms. 
  
Here, we investigate based on density functional theory (DFT) as implemented 
in the {\tt FLEUR}~\cite{FLEUR} code and in the {\tt VASP} code~\cite{vasp2} how pair-wise Heisenberg and higher-order 
exchange constants are modified at transition-metal interfaces. We consider freestanding trilayers 
and start with Rh/Fe/Ir which serves as a simplified model of the film system Rh/Fe/Ir(111). 
In our first-principles
calculations, we replace either the Rh layer by another $4d$ transition metal or the Ir layer by another $5d$ 
transition metal. Thereby, we can increase or decrease the $4d$ or $5d$ band filling which influences the hybridization with the $3d$ Fe states and via changes in the electronic structure also the magnetic interactions. 
Since a strong influence of the stacking sequence has been reported for Rh/Fe/Ir(111), we compare fcc and
hcp stacking in our theoretical study.
Finally, we replace the Fe layer by a Co layer to investigate the effect of changing the transition metal with the
intrinsic magnetic moment and to obtain results for the effect of higher-order interactions in Co.
From the freestanding trilayers we find a number of interesting trends concerning the pair-wise and higher-order 
exchange interactions at such transition-metal interfaces.
We compare these trends with the results obtained for selected ultrathin film systems which are amenable to experiments. 

The structure of the paper is as follows. First, we introduce the atomistic spin model including all relevant magnetic interactions which we calculated using DFT as well as the applied computational methods in section~\ref{Theo_Background} and section~\ref{Comp_methods}, respectively.  
The main part of the paper (section~\ref{Results_discussion}) covers both results and discussions for the investigated Fe and Co based trilayer systems separately as well as a comparison between the properties of the two of them. We relate the
freestanding trilayer calculations to those for ultrathin films on surfaces which have in the past partly been studied by
scanning tunneling microscopy (STM) experiments. Finally, we summarize our main conclusions in section~\ref{conclusion}.   

\section{Computational details}
\subsection{Atomistic spin model}
\label{Theo_Background}

In order to describe the magnetic properties of transition metal trilayers we resort to an extended version of the classical Heisenberg model dealing with magnetic moments $\mathbf{M}_i$
localized at atomic sites $i$ of a hexagonal lattice:
\begin{equation}
\label{eq:Hamiltonian}
\begin{gathered}
 \mathcal{H}=  -\sum_{ij} J_{ij}\mathbf{m}_i \cdot \mathbf{m}_j   -\sum_{ij} B_{ij}(\mathbf{m}_i \cdot \mathbf{m}_j)^2\\
 -\sum_{ijkl}K_{ijkl}[(\mathbf{m}_i \cdot \mathbf{m}_j)(\mathbf{m}_k \cdot \mathbf{m}_l)+(\mathbf{m}_i \cdot \mathbf{m}_l)(\mathbf{m}_j \cdot \mathbf{m}_k)\\
 -(\mathbf{m}_i \cdot \mathbf{m}_k)(\mathbf{m}_j \cdot \mathbf{m}_l)]\\
 -\sum_{ijk}Y_{ijk}[(\mathbf{m}_i \cdot \mathbf{m}_j)(\mathbf{m}_j \cdot \mathbf{m}_k)+(\mathbf{m}_k \cdot \mathbf{m}_j)(\mathbf{m}_j \cdot \mathbf{m}_i)]
 \end{gathered}
 \end{equation}

Here we use the normalized vector of the magnetic moment, $\mathbf{m}_i=\mathbf{M}_i/M_i$.
The first contribution to the Hamiltonian comes from the pair-wise Heisenberg exchange interaction which can be derived in second order pertubation theory from the Hubbard model for spin 1/2-particles in case the Coulomb interaction is large compared to the hopping of electrons~\cite{Hubbard63, MacDonald88}. The sign of the exchange constant $J_{ij}$ determines whether a parallel orientation, i.e.~a ferromagnetic (FM) coupling ($J>0$), or an antiparallel ordering, i.e.~an antiferromagnetic (AFM) coupling ($J<0$) between moments $\mathbf{m}_i$ and $\mathbf{m}_j$ is favored by this 
bilinear interaction.   
Different signs of the corresponding exchange constants $J_i$ result in frustration which can lead to a tilting of the magnetic moments thereby being the origin of noncollinear spin structures such as spin spirals.
\begin{figure}[htbp]
	\centering
	\includegraphics[scale=0.6,clip]{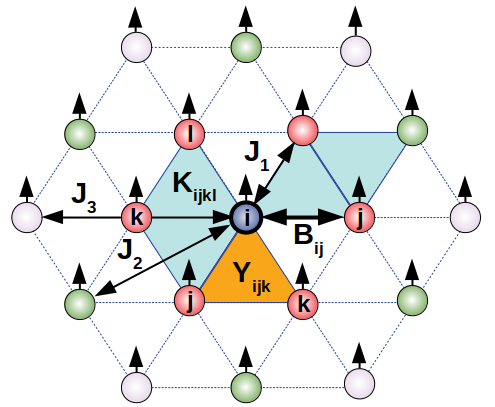}
	\caption{Schematic representation of the four considered magnetic exchange interactions on the two-dimensional hexagonal lattice. The reference atom is depicted in blue color, its nearest neighbors are marked in red, the next-nearest in green and the third nearest in light blue color.	
	Whereas both the pair-wise Heisenberg (shown up to the third-nearest neighbor) and the biquadratic exchange involve only two lattice sites, the three-site four spin interaction contains electron hopping over the three lattice sites \textit{ijk} thereby forming a triangle depicted in orange. One out of the 6 possible triangles
	is illustrated. The four-site four spin interaction on the other hand includes hopping of electrons over the four sites \textit{ijkl} resulting in a diamond shape. Two out of 12 possible diamonds for the nearest-neighbor (NN) approximation are depicted in blue color.}
	\label{HexLattice}
\end{figure}

The second part of the Hamiltonian of Eq.~(\ref{eq:Hamiltonian}) describes the biquadratic exchange, an interplay which involves four times the hopping of an electron between the two lattice sites $i$ and $j$ thereby representing a so-called two-site four spin interaction. As for the Heisenberg exchange, the biquadratic constant $B_{ij}$ is identical for
symmetry equivalent pairs of spins.
In this paper only biquadratic exchange with nearest neighbors is taken into account as shown in Fig.~\ref{HexLattice}. 

Expanding the Hubbard model up to fourth order perturbation theory in the limit of large Coulomb energy $U$ with respect
to the hopping parameter $t$~\cite{Takahashi_1977}, one finds in addition the four-site four spin interaction characterized by the parameter $K_{ijkl}$ (cf.~Eq.~(\ref{eq:Hamiltonian})). Allowing an electron to hop over four lattice sites $i$,$j$,$k$,$l$, one finds closed hopping paths taking the shape of a diamond on the hexagonal lattice within the nearest-neighbor (NN) approximation (see Fig.~\ref{HexLattice}).

The last interaction considered here and listed as the fourth contribution in Eq.~(\ref{eq:Hamiltonian}) has recently been derived from a multi-band Hubbard model~\cite{Hoffmann2020} in fourth order perturbation theory: the three-site four spin interaction. In contrast to the previously mentioned contribution to the Hamiltonian, this term leads to triangle shaped hopping paths for an electron on the hexagonal lattice which is highlighted in orange color in Fig.~\ref{HexLattice}.
\begin{figure*}[htbp]
	\centering
	\includegraphics[scale=0.5,clip]{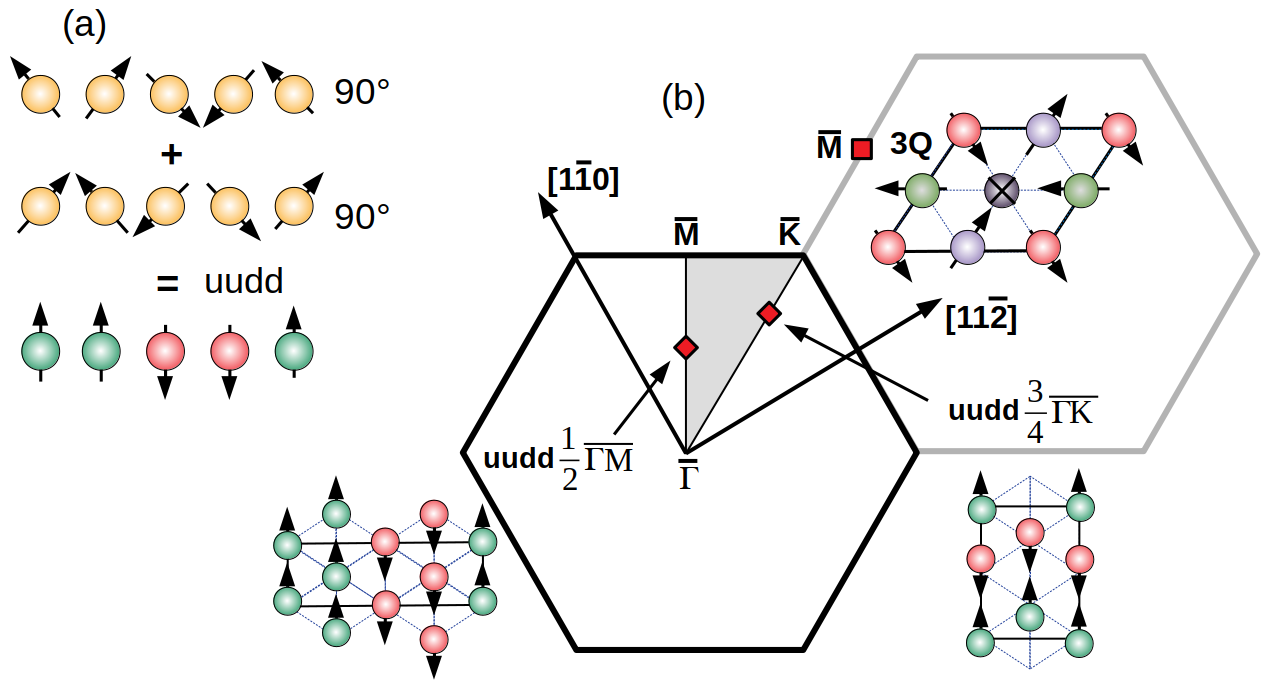}
	\caption{(a) The superposition of two  90$^{\circ}$ spin spirals with opposite rotational sense leads to the formation of a collinear \textit{uudd} state. (b) Sketch of the two-dimensional hexagonal Brillouin zone with the position of $\mathbf{q}$ vectors for 90$^{\circ}$ spin spirals forming the two possible \textit{uudd} states at $\mathbf{q}$=$\pm \frac{1}{2}\overline{\Gamma \text{M}}$ and $\mathbf{q}$=$\pm \frac{3}{4}\overline{\Gamma \text{K}}$. The collinear spin structures within their respective unit cells used for our calculations are also depicted besides the position of the three-dimensional \textit{3Q} state which results from a superposition of $\mathbf{q}$ vectors at three different $\overline{\text{M}}$-points.}
	\label{HexBZ}
\end{figure*}

In the following we denote the biquadratic, four-site four spin and three-site four spin interaction by $B_1$, $K_1$, 
and $Y_1$, respectively, in order to bear the NN approximation in mind. Since the exchange interactions are our focus 
in this work we neglect 
-- both in the Hamiltonian (Eq.~(\ref{eq:Hamiltonian})) and in our DFT calculations --
the influence of spin-orbit coupling which is the origin of the Dzyaloshinskii-Moriya interaction, the magnetocrystalline anisotropy, and recently proposed multi-spin chiral interactions \cite{Brinker2019,Brinker2020,Mankovsky2020,Laszloffy2019}. 

Now we turn our focus to the question of how these exchange constants can be calculated by means of DFT. To obtain
the pair-wise exchange constants, we use the energy dispersion of spin spirals.
Spin spirals are characterized by a spin spiral vector $\mathbf{q}$ and
describe spin structures with magnetic moments tilted by a constant angle for adjacent moments along the direction
given by $\mathbf{q}$.
Hence the moment at lattice site $\mathbf{R}_i$ is given by 
$\mathbf{M}_i = M (\cos(\mathbf{q}\cdot\mathbf{R}_i), \sin(\mathbf{q}\cdot\mathbf{R}_i),0)$,
where $M$ is the constant magnitude of the magnetic moment at all sites. 
Since spin spirals represent the general solution of the classical Heisenberg model on a periodic lattice, the
pair-wise exchange constants can be obtained from fitting energy dispersions $E(\mathbf{q})$ of flat homogeneous spin spirals along the high symmetry directions $\overline{\Gamma \text{M}}$ and $\overline{\Gamma \text{KM}}$ of the two-dimensional (2D) hexagonal Brillouin zone (BZ) (see Fig.~\ref{HexBZ}(b)). 
At the center of the 2D BZ, i.e. at~$\mathbf{q}=0$ represented by 
the $\overline{\Gamma}$-point the FM state occurs, whereas at the $\overline{\text{M}}$-point the row-wise antiferromagnetic (RW-AFM) state and at the $\overline{\text{K}}$-point the N\'eel state with angles of
$120^\circ$ between adjacent moments can be found. 
Note that we restrict the discussion of trends to Heisenberg exchange constants
up to the third nearest neighbor (Fig.~\ref{HexLattice}). However, the fitting of the energy dispersion
of spin spirals has been performed with exchange up to eight nearest neighbors.

A superposition of spin spirals -- also referred to as 1Q-states -- with symmetry equivalent $\mathbf{q}$-vectors 
can lead to the formation of multi-Q states. 
A prominent example is the collinear \textit{uudd} or double-row wise antiferromagnetic state~\cite{Hardrat2009} originating from a superposition of two counterpropagating 90$^{\circ}$ spin spirals (see Fig.~\ref{HexBZ}(a)). 
In the 2D hexagonal BZ there are two possibilities to obtain a 90$^{\circ}$ spin spiral and hence an 
\textit{uudd} state by the superposition of its left- and right-rotating compounds: at $\mathbf{q}$=$\pm \frac{1}{2}\overline{\Gamma \text{M}}$ and $\mathbf{q}$=$\pm \frac{3}{4}\overline{\Gamma \text{K}}$ (Fig.~\ref{HexBZ} (b)). 
A third multi-Q state needed for the evaluation of the HOI parameters is the \textit{3Q} state~\cite{Kurz2001}, 
a three-dimensional noncollinear spin structure on the 2D lattice formed by a linear combination of three orthogonal symmetry equivalent 1Q-states at the $\overline{\text{M}}$-point of the hexagonal BZ (Fig.~\ref{HexBZ}(b)). 

Within the classical Heisenberg model these three multi-Q states are energetically degenerate with their corresponding 
1Q spin spiral states. This degeneracy is lifted by the HOI terms which can couple the single 1Q-modes to form the
superposition states. Keeping this fact in mind, one can calculate the exchange parameters beyond pair-wise
Heisenberg exchange
within the NN approximation by solving a system of three coupled equations~\cite{Hoffmann2020}:  
\begin{eqnarray}
   \Delta E_{\overline{\text{M}}} & = & E_{\text{3Q}}-E_{\overline{\text{M}},\text{1Q}} = \frac{16}{3}(2K_1 + B_1 - Y_1)
	\label{eq:Delta_E_3Q} \\
   \Delta E_{\frac{1}{2}\overline{\Gamma\text{M}}} & = & E_{\text{uudd},\frac{\overline{\text{M}}}{2}} - E_{\frac{\overline{\text{M}}}{2},\text{1Q}} = 4(2K_1 - B_1 - Y_1) \label{eq:Delta_E_GM} \\
   \Delta E_{\frac{3}{4}\overline{\Gamma\text{K}}} & = & E_{\text{uudd},\frac{3\overline{\text{K}}}{4}} - E_{\frac{\overline{3\text{K}}}{4},\text{1Q}} = 4(2K_1 - B_1 + Y_1) \label{eq:Delta_E_GK}
\end{eqnarray}
The respective four-atomic unit cells per layer from which both the total energies of the \textit{uudd} states, $E_{\text{uudd},\frac{\overline{\text{M}}}{2}}$ and $E_{\text{uudd},\frac{3\overline{\text{K}}}{4}}$, as well as the energy of the 
\textit{3Q} state, $E_{\text{3Q}}$, can be calculated via DFT are sketched in Fig.~\ref{HexBZ}(b) as insets. Here, $E_{\overline{\text{M}},\text{1Q}}$, $E_{\frac{\overline{\text{M}}}{2},\text{1Q}}$ and $E_{\frac{\overline{3\text{K}}}{4},\text{1Q}}$ denote the respective total energy of the corresponding single-Q state. Inverting Eq.~(\ref{eq:Delta_E_3Q}$-$\ref{eq:Delta_E_GK}) yields a 
direct way of calculating the NN HOI terms according to:
\begin{eqnarray}
   B_1 & = & \frac{3}{32} \Delta E_{\overline{\text{M}}}-\frac{1}{8}\Delta E_{\frac{1}{2}\overline{\Gamma\text{M}}} 
	\label{eq:B1}\\
   Y_1 & = & \frac{1}{8}(\Delta E_{\frac{3}{4}\overline{\Gamma\text{K}}}-\Delta E_{\frac{1}{2}\overline{\Gamma\text{M}}})
	\label{eq:Y1}\\
   K_1 & = & \frac{3}{64}\Delta E_{\overline{\text{M}}}+\frac{1}{16}\Delta E_{\frac{3}{4}\overline{\Gamma\text{K}}}
	\label{eq:K1}
\end{eqnarray}
At this point it must be mentioned that these parameters alter the first three Heisenberg exchange constants obtained from fitting the DFT energy dispersion of spin spirals 
since their energy contribution to $E(\mathbf{q})$ has the same functional form, 
e.g.~for instance~$B_1$ and $J_3$ cannot be specified separately from each other. 
Typically, the $J_{i}$ are obtained by fitting the DFT spin spiral
energy dispersions by only the first term of Eq.~(\ref{eq:Hamiltonian}), i.e.~only
the Heisenberg exchange.
If the full spin model given by Eq.~(\ref{eq:Hamiltonian}) is to be applied, i.e.~if the
higher-order terms are included, then the first three of the $J_{i}$ need to be
replaced by the modified values $J_{\text{i}}^\prime$ given by~\cite{Paul2020}:
\begin{eqnarray}
   J^{\prime}_{\text{1}} & = & J_1-Y_1 \label{eq:J1_prime}\\
   J^{\prime}_{\text{2}} & = & J_2-Y_1 \label{eq:J2_prime}\\
   J^{\prime}_{\text{3}} & = & J_3-\frac{1}{2}B_1 \label{eq:J3_prime} \text{.}
\end{eqnarray}
In contrast to $Y_1$ and $B_1$, the four-site four spin term $K_1$ yields a constant value of $-12K_1$ for all spin spirals
independent of $\mathbf{q}$ and hence does not influence the obtained exchange constants. 
The modification according to Eqs.~(\ref{eq:J1_prime}-\ref{eq:J3_prime}) has to be taken into account for further simulations based on
the atomistic spin model such as spin dynamics 
or Monte-Carlo simulations.

\subsection{Computational methods}
\label{Comp_methods}
We have performed DFT calculations within the framework of the all-electron full-potential linearized augmented planewave 
(FLAPW) method~\cite{Kurz2000,Hamann1979,Wimmer1981} as implemented in the {\tt FLEUR} code~\cite{FLEUR} and the projector augmented wave (PAW) method~\cite{Bloechl1994,Kresse1999} based on the pseudopotential approach as incorporated in the 
{\tt VASP} code~\cite{vasp2,Kresse1996}. We used the local density approximation (LDA) with the interpolation proposed by Vosko, Wilk and Nusair (VWN)~\cite{Vosko1980} to include exchange-correlation effects. For all trilayer and film systems 
we have chosen the theoretical LDA in-plane lattice constant of Ir, 
i.e.~$d_{\rm Ir}=$2.70 {\AA}~\cite{Dupe2014}, since the starting point of our investigation is the film system 
Rh/Fe/Ir(111) studied previously~\cite{Romming2018}. In order to obtain
trends from our trilayer calculations which reflect only the change of the transition-metal elements we also kept the
interlayer distances fixed and as determined for Rh/Fe/Ir(111)
(see Table~\ref{tab:interlayerdistance}). 

For the trilayers we only change the $4d$ or $5d$ element, while the geometric structure is kept fixed. 
Besides Rh as a $4d$ transition metal we consider Tc, Ru, and Pd along the row of the periodic table, whereas the $5d$ element Ir is replaced by Re, Os, and Pt. In a similar fashion the
relaxed interlayer distances from Rh/Co/Ir(111)~\cite{Meyer2019} serve as a starting point to calculate the exchange parameters for Co based trilayers. In addition, the properties of symmetric $4d$/Fe/$4d$ systems have been evaluated based on the in-plane lattice constant of Rh which amounts to 2.72 {\AA} and the unrelaxed interlayer distances of bulk Rh 
(see Table~\ref{tab:interlayerdistance}).
\begin{table}[htb]
	\centering
	\caption{Overview of the in-plane lattice constants and the relaxed interlayer distances from the respective film calculations for all trilayer systems investigated in this paper. The theoretical in-plane lattice constant of Ir is taken from Ref.~\cite{Dupe2014}, the in-plane lattice constant of Rh from Ref.~\cite{Meyer2017_2}, relaxed interlayer distances for Fe based trilayers from Ref.~\cite{Romming2018} and for Co based trilayers from Ref.~\cite{Meyer2019}. For symmetric $4d$/Fe/$4d$ systems the unrelaxed interlayer distance of Rh bulk has been used. All values are given in {\AA}.}
	\label{tab:interlayerdistance}
	\begin{ruledtabular}
		\begin{tabular}{l c c c }
			& d$_{\text{Ir}}$ & d$_{4d\text{-Fe}}$  & d$_{\text{Fe-Ir}}$    \\ 
			\colrule
			fcc-$4d$/Fe/Ir & 2.70  & 1.97  & 2.07  \\
			hcp-$4d$/Fe/Ir & 2.70 & 1.97  & 2.06  \\
			\colrule
			& d$_{\text{Ir}}$ & d$_{5d\text{-Fe}}$  & d$_{\text{Fe-Rh}}$    \\ 
			\colrule
			fcc-$5d$/Fe/Rh & 2.70  & 1.97  & 2.07  \\
			\colrule
			& d$_{\text{Rh}}$  & d$_{4d\text{-Fe}}$ & d$_{\text{Fe-}4d}$   \\ 
			\colrule
			fcc-$4d$/Fe/$4d$ & 2.72 & 2.22  & 2.22  \\
			\colrule
			& d$_{\text{Ir}}$ & d$_{4d\text{-Co}}$ & d$_{\text{Co-Ir}}$ \\
			\colrule
			fcc-$4d$/Co/Ir &2.70 & 2.06 & 2.15  \\
			 \end{tabular} 
	\end{ruledtabular}
\end{table}

The spin spiral energy dispersions and total energies of the collinear \textit{uudd} states have been calculated
via the FLAPW method for all Fe and Co based trilayers. 
Using the generalized Bloch Theorem~\cite{Kurz2000} total energies of flat spin spirals can be computed self-consistently 
in the chemical unit cell
within the scalar-relativistic approximation. We have used 1936 $k$-points in the full 2D BZ and a large cutoff of
$k_{\text{max}}=4.1$~a.u.~$^{-1}$ to ensure convergence with the basis function set. The muffin tin radii have been chosen as 2.31 a.u.~for both $4d$ and $5d$ elements, while a slightly smaller value of 2.23 a.u.~has been set for the magnetic elements Fe and Co. For the $uudd$ states only the number of $k$-points has been adapted in such a way that its density in the full 2D BZ corresponds approximately to that used for the respective spin spiral calculation; hence, the $k$-point set amounts to 168 in the irreducible part of the first BZ for the $uudd$ state along $\overline{\Gamma \text{M}}$ and to 336 for the $uudd$ state along $\overline{\Gamma \text{K}}$ direction. 

For Co based trilayers the energy differences between the \textit{3Q} and the RW-AFM state have been calculated by the FLAPW method within the four-atomic unit cell per layer as well; here, 484 $k$-points in the full 2D BZ have been chosen which is exactly one quarter of the number of $k$-points used for the respective spin spiral calculation.

For the Fe based trilayers (and the Rh/Fe/Rh(111) film system) the energy differences between the \textit{3Q} and the RW-AFM state have been calculated using the PAW method. A $28\times28\times1$ $\overline{\Gamma}$-centered ($15\times15\times1$) $k$-point mesh and the standard energy cutoffs as supplied by the respective potential file library of
{\tt VASP} have been used. In Appendix A we demonstrate that mixing both DFT methods in this way is a 
valid procedure for obtaining trends of HOI parameters in Fe based transition-metal trilayers and film systems by 
applying convergence tests to selected systems. 

We have also determined the HOI parameters for selected film systems. As discussed in the previous section, we
need the energy dispersion $E(\mathbf{q})$ of spin spirals as well as the total energies of the three multi-Q states. 
For the film systems Pd/Fe/Ir(111), Rh/Fe/Ir(111), and Rh/Co/Ir(111) we have used $E(\mathbf{q})$ from our previous 
work \cite{Dupe2014,Romming2018,Meyer2019}. For these systems we have calculated for both stackings of the Pd or
Rh overlayer via {\tt FLEUR} the missing 
multi-Q states with the same setup of the films as in the cited references. For Pd/Fe/Ir(111) and Rh/Co/Ir(111), we 
have calculated the \textit{3Q} and the two $uudd$ states, while only the \textit{3Q} state was required for Rh/Fe/Ir(111), 
since both $uudd$ states have been calculated previously~ \cite{Romming2018}.
For fcc- and hcp-Rh/Fe/Rh(111), we have calculated $E(\mathbf{q})$ and the two $uudd$ states using {\tt FLEUR}. The 
total energy
difference between the RW-AFM and the \textit{3Q} state has been obtained using {\tt VASP}. Asymmetric films with 9 Rh layers
have been used for all calculations and the film structure was relaxed.
For fcc-Pd/Co/Ir(111), we have calculated $E(\mathbf{q})$ and all three multi-Q states using {\tt FLEUR}. Asymmetric 
films with 9 Ir layers have been used for all calculations and the film structure was relaxed \cite{MeyerPhD}.
For all multi-Q state calculations we used the same k-point meshes as for the freestanding trilayers and
a value of $k_{\rm max}=4.1$~a.u.$^{-1}$.

\section{Results and discussion}
\label{Results_discussion}
\subsection{Fe based trilayers}
\label{Fe_trilayers}

\begin{figure}[htbp]
	\centering
	\includegraphics[scale=0.48,clip]{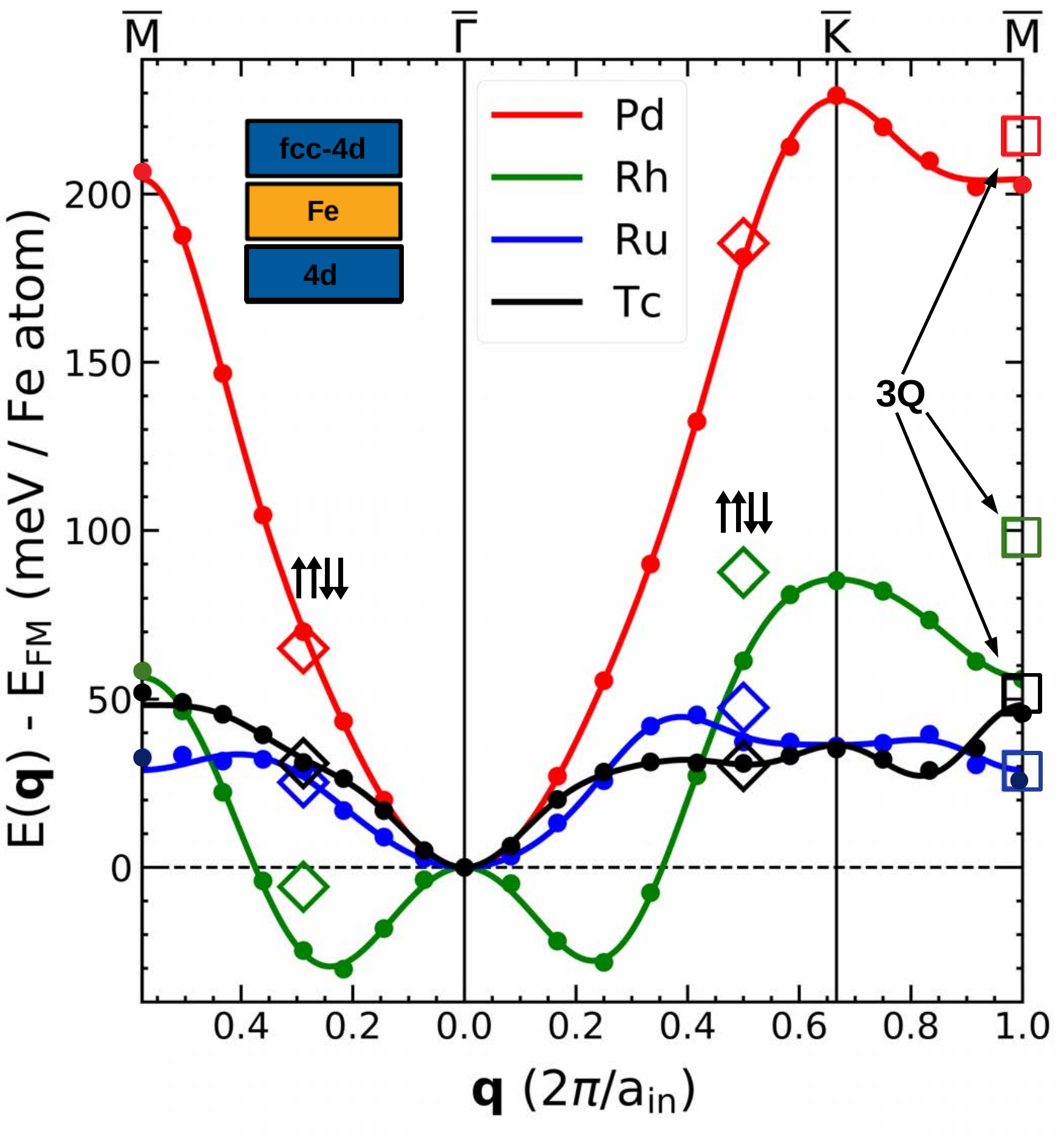}
	\caption{Energy dispersion of flat spin spirals for symmetric fcc-$4d$/Fe/$4d$ trilayers. Total energies calculated by DFT are marked by filled circles, whereas the solid lines represent fits to the Heisenberg model. Both energies of the \textit{uudd} ($\uparrow \uparrow \downarrow \downarrow$) states as well as the \textit{3Q} state are plotted as empty diamonds 
	and squares at the \textit{q} values of the corresponding single-Q states, respectively.   }
	\label{SS_4dFe4d}
\end{figure}

We start our discussion with the trend of Heisenberg exchange and HOI parameters for symmetric $4d$/Fe/$4d$ trilayers. 
In these systems, the effect of the band filling in the $4d$ layer on the magnetic properties of the Fe layer is already clearly revealed. 
Note that our trilayers are element-wise and structurally symmetric since we use the same $4d$-Fe distance for both interfaces (see Table~\ref{tab:interlayerdistance}).
These results also facilitate to understand the trends for the more complex asymmetric trilayers 
$4d$/Fe/Ir and $5d$/Fe/Rh. Note that these trilayers are isoelectronic if we use the $4d$ and $5d$ transition metal
with the same number of $d$ electrons.
The reference system for these trilayers is Rh/Fe/Ir. For the corresponding ultrathin film system Rh/Fe/Ir(111) a
stacking-dependent change of the magnetic ground state has been observed \cite{Romming2018}. Therefore, we consider
both fcc- and hcp-stacking of $4d$/Fe/Ir trilayers. Finally, we compare the trends from the trilayers with experimental
and theoretical results for the corresponding ultrathin film systems. 

\begin{figure*}[htbp]
	\centering
	\includegraphics[scale=0.4,clip]{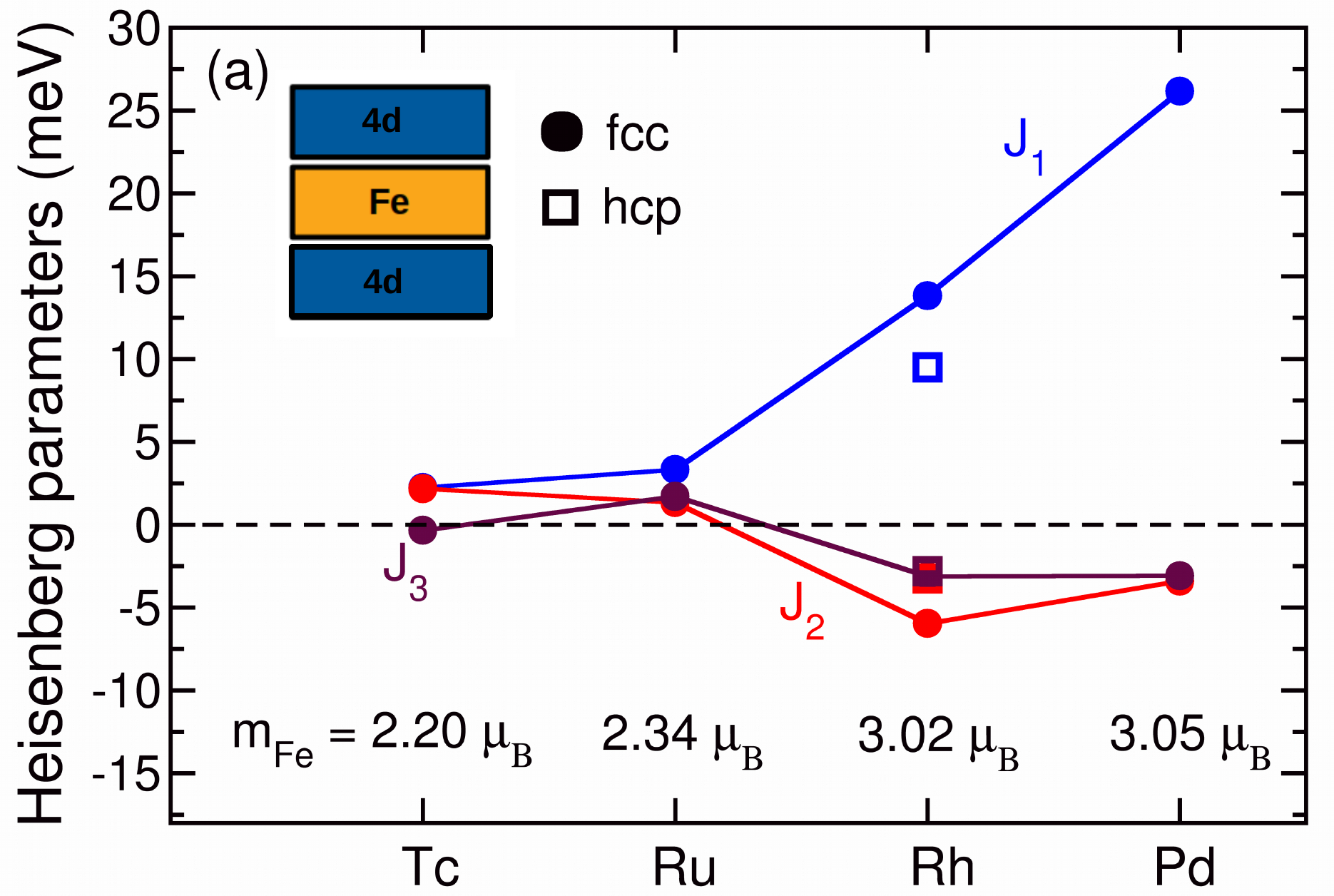}
	\includegraphics[scale=0.4,clip]{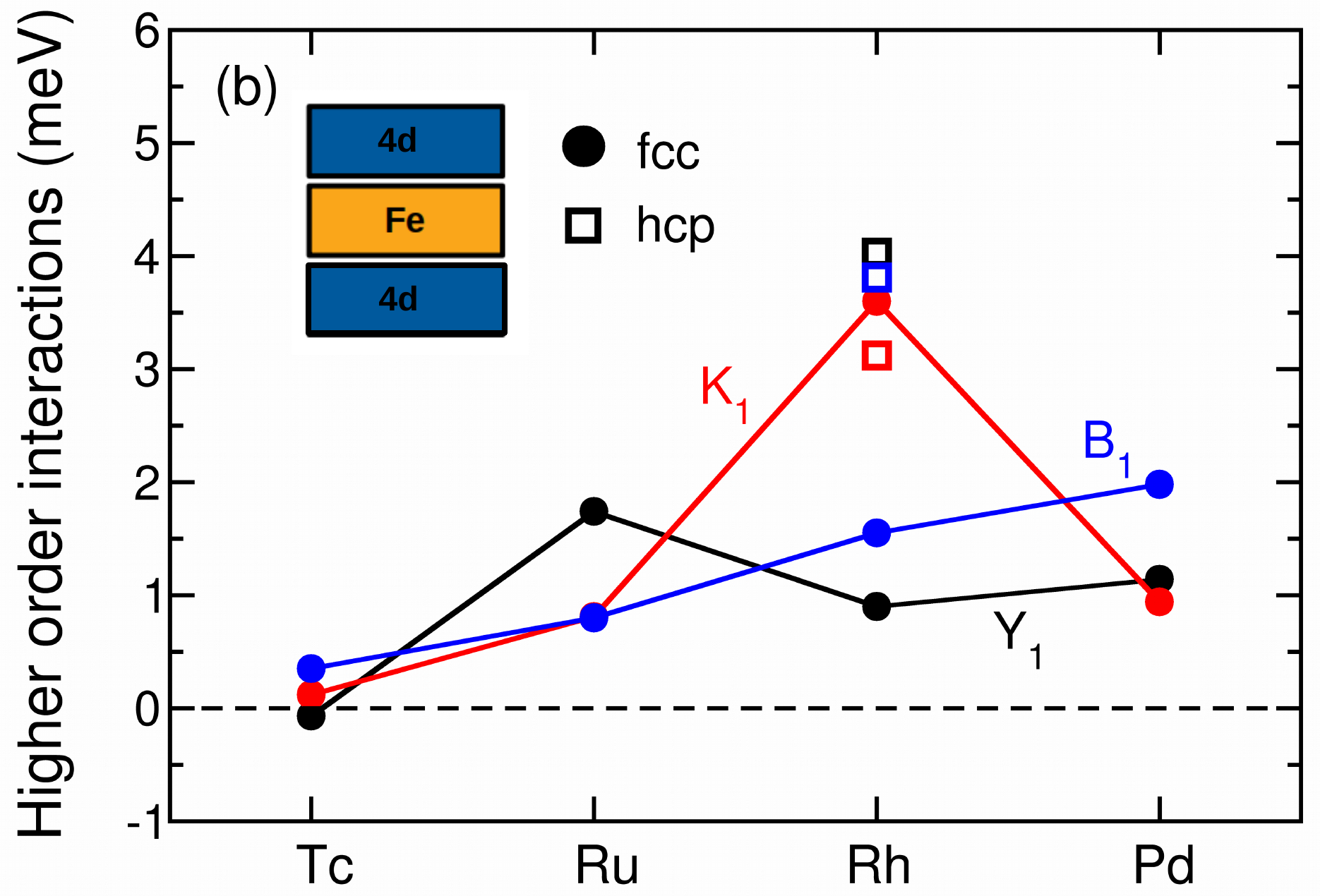}
	\caption{Comparison of the trend for (a) the Heisenberg exchange constants as extracted from fitting the corresponding spin spiral dispersions, i.e. without modification by HOI terms, and (b) the HOI parameters for symmetric $4d$/Fe/$4d$ trilayers. The lines connecting the data points serve as a guide to the eye. }
	\label{J__HOI_4dFe4d}
\end{figure*}

\textbf{Symmetric $\bm{4d}$/Fe/$\bm{4d}$ trilayers.} We 
begin
with the energy dispersion $E(\mathbf{q})$ of flat spin spirals in 
symmetric $4d$/Fe/$4d$ trilayers in fcc stacking
along both high symmetry directions of the hexagonal 2D BZ (Fig.~\ref{SS_4dFe4d}). 
The energy dispersion of Pd/Fe/Pd is characteristic for a system dominated by a nearest-neighbor ferromagnetic 
exchange interaction. The FM state at the $\overline{\Gamma}$-point exhibits the lowest energy and the dispersion
rises quickly for spin spirals with increasing $\mathbf{q}$, i.e.~decreasing spin spiral period. The RW-AFM state 
at the $\overline{{\rm M}}$-point and the N\'eel state at the $\overline{{\rm K}}$-point are more than 200~meV/Fe atom higher
than the FM state. For Rh/Fe/Rh, the energy dispersion is qualitatively 
different: 
the RW-AFM state and the
N\'eel state have decreased in energy and spin spiral energy minima occur along both high symmetry directions. 
The energy dispersions of Ru/Fe/Ru and Tc/Fe/Tc look similar with the FM state being lowest but small energy differences
with respect to the states at the zone boundaries. 

In Fig.~\ref{SS_4dFe4d} the energies of the three multi-Q states with respect to the FM state are shown at the $\mathbf{q}$ vectors of the corresponding single-Q states.
For Fe trilayers with Pd, Ru, and Tc overlayers, the multi-Q states
are energetically relatively close to the spin spiral energy dispersion. However, 
for Rh/Fe/Rh the multi-Q states are 
quite far from the respective spin spiral energies, by about 20 to 40 meV/Fe atom, indicating a significant 
contribution of higher-order exchange interactions. However, in none of the fcc-$4d$/Fe/$4d$ 
trilayers a multi-Q state is the energetically lowest magnetic structure.

\begin{table}[htb]
	\centering
	\caption{Magnetic moments of the symmetric fcc-$4d$/Fe/$4d$ trilayers for the ferromagnetic state. All values are given in $\mu_B$.}
	\label{table:4dFe4d}
	\begin{ruledtabular}
		\begin{tabular}{l c c c c }
			$4d$/Fe/$4d$& $m_{4d}$&$m_{\text{Fe}}$ & $m_{4d}$ \\
			\colrule
			Tc/Fe/Tc&$-0.08$ &$+2.20$ &$-0.08$ \\
			Ru/Fe/Ru&$-0.35$ &$+2.34$ &$-0.35$ \\
			Rh/Fe/Rh&$+0.33$ &$+3.02$ &$+0.33$ \\
			Pd/Fe/Pd&$+0.30$ &$+3.05$ &$+0.30$ \\
			 \end{tabular} 
	\end{ruledtabular}
\end{table}	

The magnetic moments of both the Fe layer and the adjacent $4d$ layers are listed in Table~\ref{table:4dFe4d}. 
Since the geometric structure is the same for all trilayers, the change of the magnetic moment in the Fe layer
arises only due to the $3d$-$4d$ hybridization. Two effects play a role. First, the $4d$ band in the 
adjacent layers moves to lower energies which modifies the hybridization with the Fe $3d$ bands. Second, 
the extent of the $4d$ orbitals decreases with increasing nuclear number due to the imperfect screening of the nuclear
Coulomb potential by the $4d$ electrons. This leads to a decrease of overlap with the $3d$ Fe states. As a result 
of both effects the Fe magnetic moment
rises with $4d$ band filling. The induced magnetic moments in the adjacent $4d$ layers can be quite considerable 
due to the large spin susceptibility of the $4d$ transition metals which also leads to the AFM alignment at the 
beginning of the series, for Tc and Ru, and FM ordering at the end of the series, for Rh and Pd.

From fitting the DFT energy dispersions of spin spirals shown in Fig.~\ref{SS_4dFe4d} the pair-wise Heisenberg exchange
constants are obtained. 
Fig.~\ref{J__HOI_4dFe4d}(a) shows the trend of the exchange constants for fcc-$4d$/Fe/$4d$ trilayers
up to the third nearest neighbor without modification by HOI terms.
The NN exchange constant, $J_1$, rises nearly linearly from Ru/Fe/Ru to Pd/Fe/Pd in accordance with the increasing
energy difference between the FM state and the RW-AFM state seen in Fig.~\ref{SS_4dFe4d}. For Tc/Fe/Tc and Ru/Fe/Ru
all exchange constants are similar in agreement with the similar energy dispersions. 
In addition, $J_2$ and $J_3$ of Tc/Fe/Tc and Ru/Fe/Ru are on the same order of magnitude as $J_1$ 
leading to a remarkable flattening of their spin spiral curves as compared to Pd/Fe/Pd. 
From Ru/Fe/Ru to Rh/Fe/Rh 
a change of sign of $J_2$ and $J_3$, i.e.~prefered AFM coupling, is observed which leads to exchange frustration 
in the latter trilayer and the spin spiral minima seen in Fig.~\ref{SS_4dFe4d}. We have also calculated spin spiral 
energy dispersions for the hcp stacking of Rh/Fe/Rh (not shown). As seen in Fig.~\ref{J__HOI_4dFe4d}(a), $J_1$ decreases 
by about 4~meV and $J_2$ is reduced by a factor of two. This is due to an energy dispersion with only shallow spin spiral
minima. 

The trend of the exchange interactions in the trilayers is consistent with DFT calculations for Fe monolayers (MLs) on 
fcc(111) and hcp(0001) surfaces of $4d$ and $5d$ transition metals~\cite{Hardrat2009}. In that work, it has been 
reported that there is a transition from AFM coupling for Fe MLs in hcp stacking on Tc(0001) and Re(0001) to FM 
coupling for Fe MLs in fcc stacking on Pd(111) and Pt(111), i.e.~an increase of $J_1$ with $d$ band filling.
Due to the fcc-stacking our trilayers do not show AFM nearest-neighbor coupling at the beginning of the series.
Nevertheless, the same explanation for the behavior of $J_1$ holds as reported in Ref.~\cite{Hardrat2009}: the observed trend is a result of the hybridization between the $3d$ bands of Fe and the $4d$ bands of the non-magnetic adjacent layers which is changed with increasing filling of electrons. This altering $3d$-$4d$ hybridization becomes also apparent in the increasing magnetic 
Fe moments along the investigated series (see Fig.~\ref{J__HOI_4dFe4d}(a)).

The HOI constants for the fcc-$4d$/Fe/$4d$ trilayers are shown in Fig.~\ref{J__HOI_4dFe4d}(b). It is apparent that
the biquadratic term rises with increasing $4d$ band filling as the NN exchange constant. In contrast,
the trend of the four-site four spin interaction displays a striking maximum for Rh/Fe/Rh trilayers. 
The three-site four spin interaction
is of similar magnitude as the biquadratic term and follows a trend more similar to the third-nearest neighbor exchange.
The HOI constants are smaller than $J_1$ in all cases, however, they can be of similar or even larger value compared
to the second and third-nearest neighbor exchange. For Ru/Fe/Ru the values of the pair-wise and higher-order exchange constants are quite comparable.

For an hcp stacking of the Rh/Fe/Rh trilayer, the biquadratic and three-site four spin interactions are drastically
enhanced to values of about 5.9 and 5.1~meV, respectively, and even exceed the value of $K_1$
(Fig.~\ref{J__HOI_4dFe4d}(b)). 
$K_1$ is slightly smaller by 0.7~meV than in the fcc case. 
Hence, there is a significant effect of the stacking order on the higher-order exchange interactions while the
pair-wise exchange interactions are much less sensitive to the stacking sequence.
As a result the $uudd$ state along $\overline{\Gamma {\rm M}}$ direction is the magnetic ground state
of the hcp-Rh/Fe/Rh trilayer. We find a change of the magnetic ground state from a spin spiral for fcc Rh
stacking to the $uudd$ state for hcp stacking also for the corresponding film system Rh/Fe/Rh(111) 
(HOI values are given in table \ref{tab:table4a}).

\begin{figure*}[htbp]
	\centering
	\includegraphics[scale=0.48,clip]{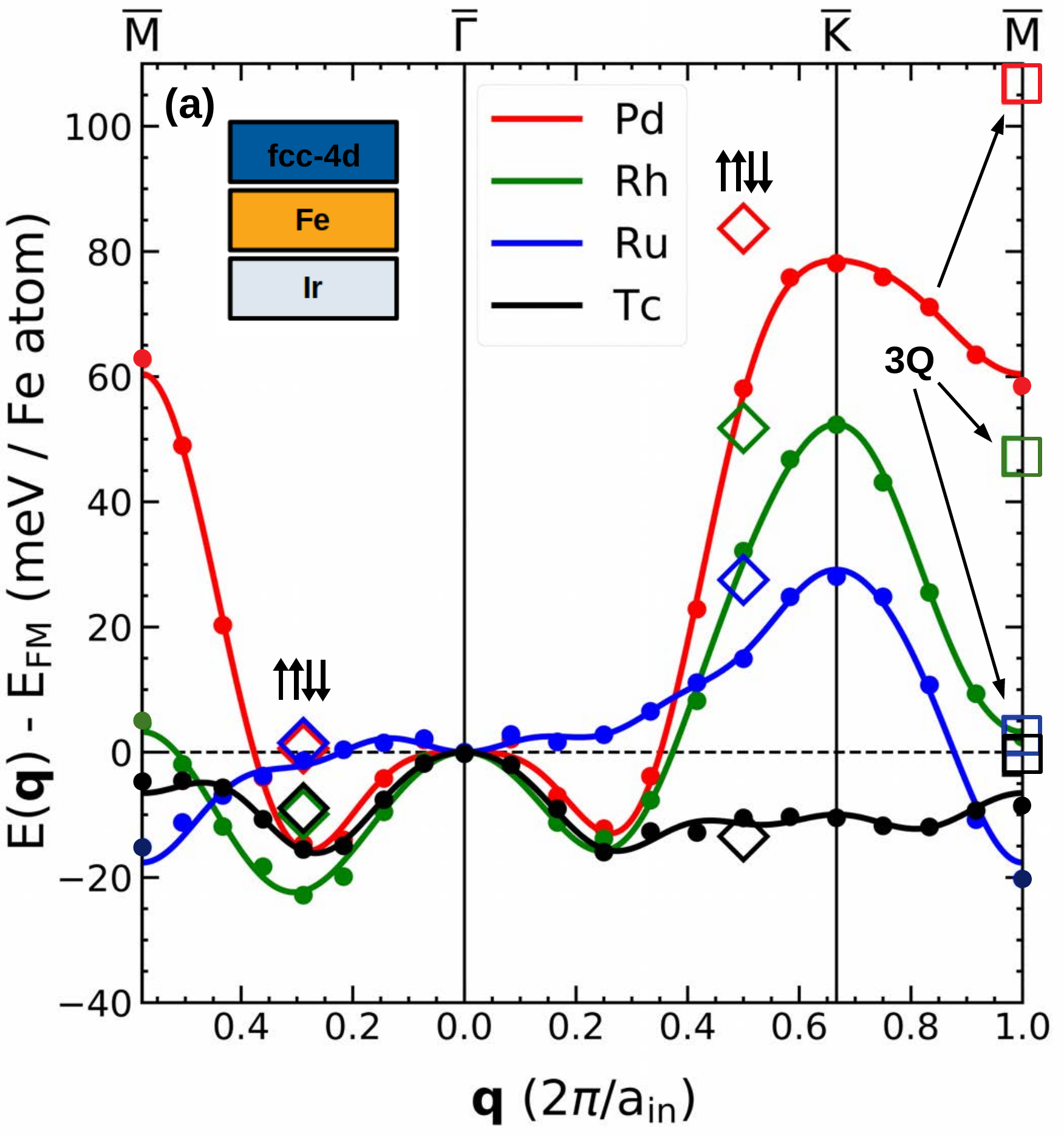}
	\includegraphics[scale=0.48,clip]{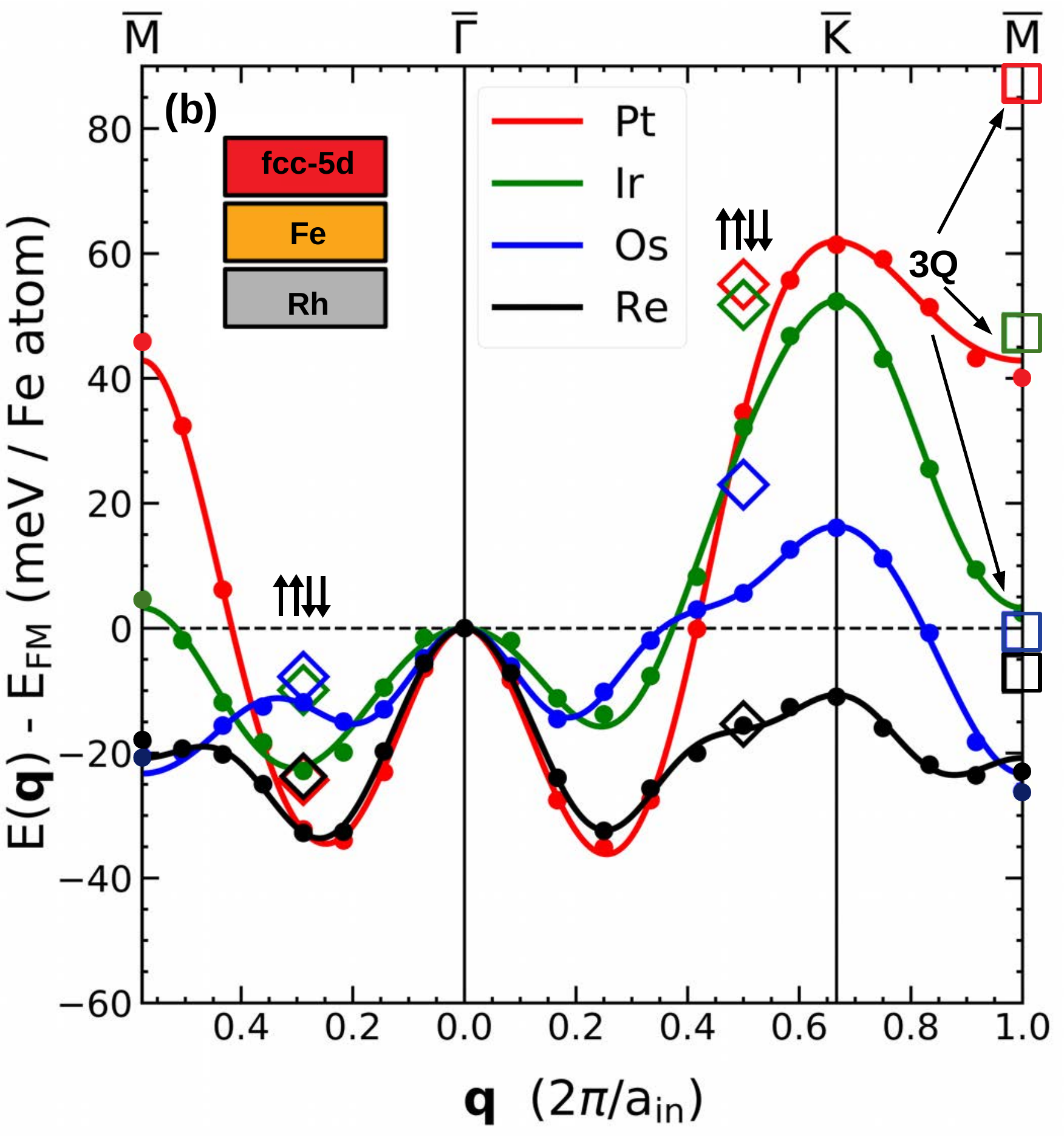}
	\caption{Energy dispersion of flat spin spirals for (a) fcc-$4d$/Fe/Ir and (b) fcc-$5d$/Fe/Rh trilayers. Total energies calculated by DFT are marked by filled circles, whereas the solid lines represent fits to the Heisenberg model. Both energies of the \textit{uudd} ($\uparrow \uparrow \downarrow \downarrow$) states as well as the \textit{3Q} state are plotted as empty diamonds and squares at the \textit{q} values of the corresponding single-Q states, respectively.   }
	\label{SS_4dFeIr_RhFe5d}
\end{figure*}

\textbf{$\bm{4d}$/Fe/Ir and $\bm{5d}$/Fe/Rh trilayers.} 
Now we turn to asymmetric Fe based trilayers. Since the starting point of our study is the film system Rh/Fe/Ir(111),
we consider $4d$/Fe/Ir trilayers. In order to investigate the effect of isoelectronic $4d$ vs.~$5d$ transition metals,
we have further performed calculations for $5d$/Fe/Rh trilayers. Fig.~\ref{SS_4dFeIr_RhFe5d} shows the energy dispersions of flat spin spirals for these two types of trilayers.
The most striking feature for all systems except those with Ru and Os are the deep energy minima along both paths of the 
BZ. The values of $\mathbf{q}$ for these minima correspond to a period of about 
$\lambda \approx$ 1.03 nm.
They are close to those found in the energy dispersion of the symmetric 
Rh/Fe/Rh trilayer (cf.~Fig.~\ref{SS_4dFe4d}). This is a result of the strong effect of the hybridization at the Fe/Rh
or the isoelectronic Fe/Ir interface which clearly dominates the energy dispersion. 

For the $4d$/Fe/Ir trilayers the
spin spiral minima are relatively similar in depths up to 23~meV/Fe atom below the FM state, while for $5d$/Fe/Rh 
trilayers the minima are deepest with a maximum value of about 35~meV/Fe atom below the FM state for the case of a 
Re and Pt overlayer. For Ru/Fe/Ir and Os/Fe/Rh, the row-wise antiferromagnetic state ($\overline{\text{M}}$ point) is the ground state, 
while all other trilayers possess a spin spiral energy minimum.
Both types of ground states indicate competing exchange interactions.
As it can be seen from Fig.~\ref{SS_4dFeIr_RhFe5d}, the $uudd$ states and the \textit{3Q} state are higher in energy than the corresponding single-Q (spin spiral) states with the only exception of the $uudd$ state along 
$\overline{\Gamma \text{K}}$ of Tc/Fe/Ir. 

The Fe atoms in all trilayers
exhibit a nearly constant magnetic moment independent of $\mathbf{q}$. Our DFT calculations reveal only small deviations 
of about 6\% from the magnetic moments of the FM state (see also Table~\ref{tab:table3}). Hence they are, to a good approximation, independent of the spin state and fulfill the basic condition of the Heisenberg model. In contrast, the moments of Ru/Fe/Ir and its isoelectronic counterpart Os/Fe/Rh show differences of up to 17\% from the FM state which necessitates the inclusion of the Stoner energy term in order to fit the DFT data (Fig.~\ref{SS_4dFeIr_RhFe5d}). Therefore, the calculated energies of these two trilayers have been shifted by $\frac{1}{2}I(M(q)-M(0))^2$ with $I$ being the Stoner parameter (420 meV for Fe~\cite{Sigalas1994}) and $M(0)$ denoting the magnetic moment of Fe at the 
$\overline{\Gamma}$-point. 

\begin{figure*}[htbp]
	\centering
	\includegraphics[scale=0.5,clip]{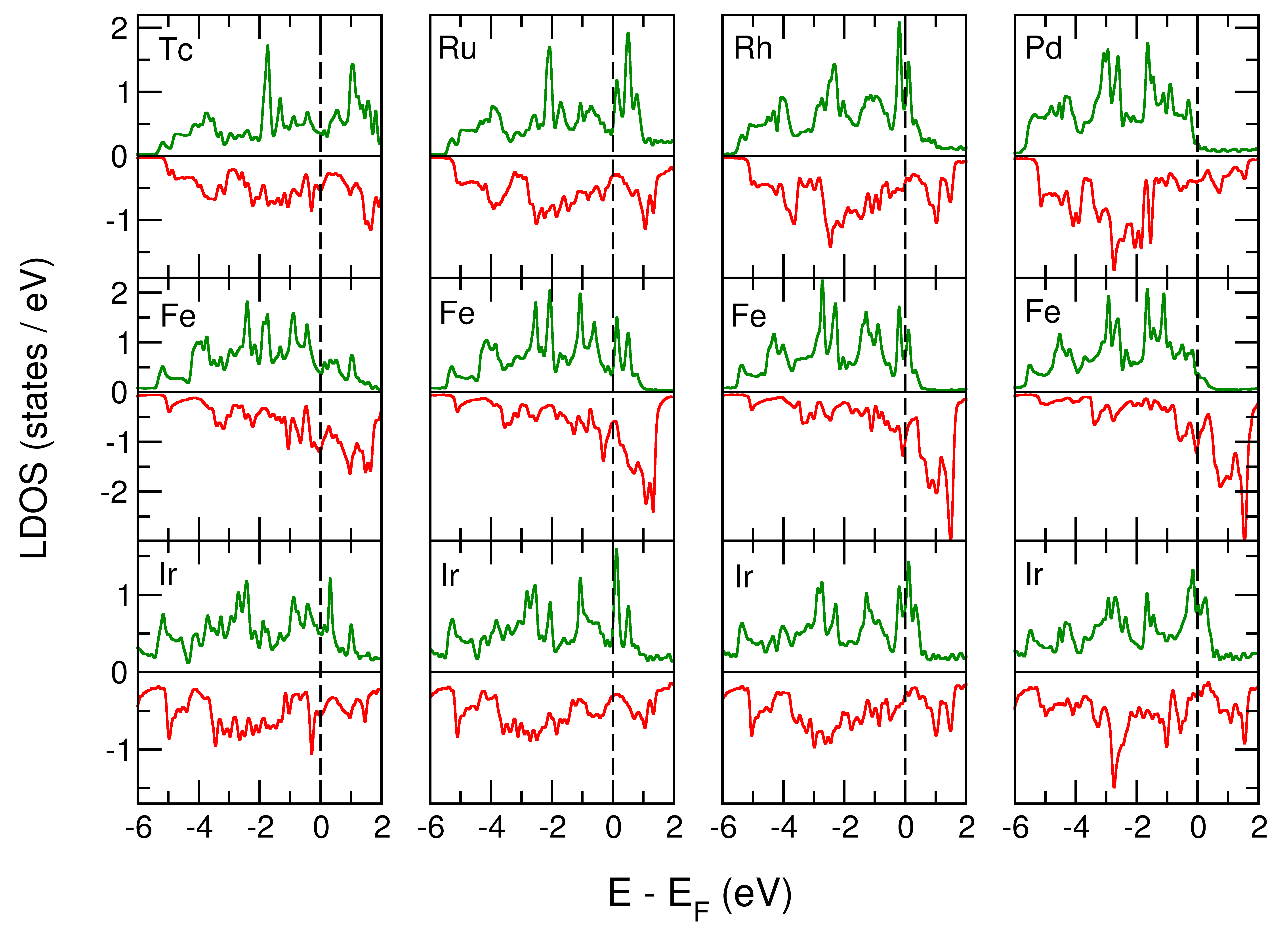}
	\caption{Local density of states (LDOS) of the fcc-$4d$/Fe/Ir trilayers. The upper panels show the LDOS of the
	$4d$ overlayer, the middle panels the LDOS of the Fe layer, and the lower panels that of the Ir layer. Majority
	and minority spin channels are represented in green and red color, respectively.}
	\label{LDOS_4dFeIr}
\end{figure*}

The varied hybridization due to the band filling of the overlayer can be seen in the local density of states shown 
for the example of $4d$/Fe/Ir trilayers in Fig.~\ref{LDOS_4dFeIr}. As we move through the $4d$ series from Tc
to Pd, the Fermi energy moves from the center to the end of the $4d$ bands (upper panels of Fig.~\ref{LDOS_4dFeIr}).
This shift of the $4d$ band affects strongly the $3d$ LDOS of the Fe layer (middle panels of Fig.~\ref{LDOS_4dFeIr}). 
In particular, the majority spin channel displays large changes in the vicinity of the Fermi energy. Thereby, the
exchange interactions are significantly modified which is manifested in the spin spiral energy dispersions discussed
above. The reduced magnetic moment of the Fe layer for trilayers (cf.~Table~\ref{tab:table3}) with an overlayer from 
the beginning of the $4d$ series can also be understood based on the variation of the majority spin LDOS. A similar 
effect as shown for the $4d$/Fe/Ir trilayers in Fig.~\ref{LDOS_4dFeIr} is found for the $5d$/Fe/Rh and the $4d$/Fe/$4d$
trilayers (not shown).

\begin{figure*}[htbp]
	\centering
	\includegraphics[scale=0.3,clip]{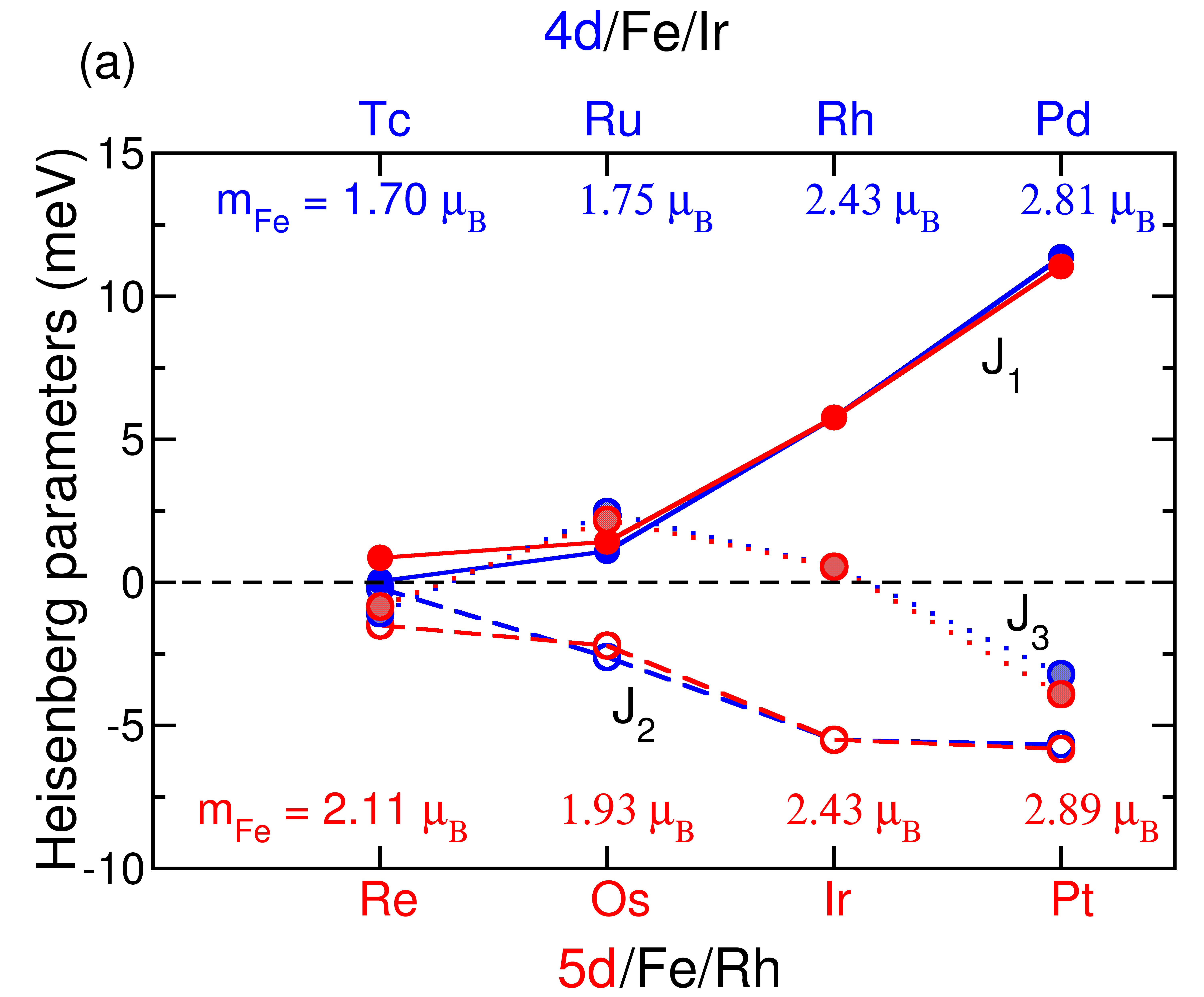}
	\includegraphics[scale=0.3,clip]{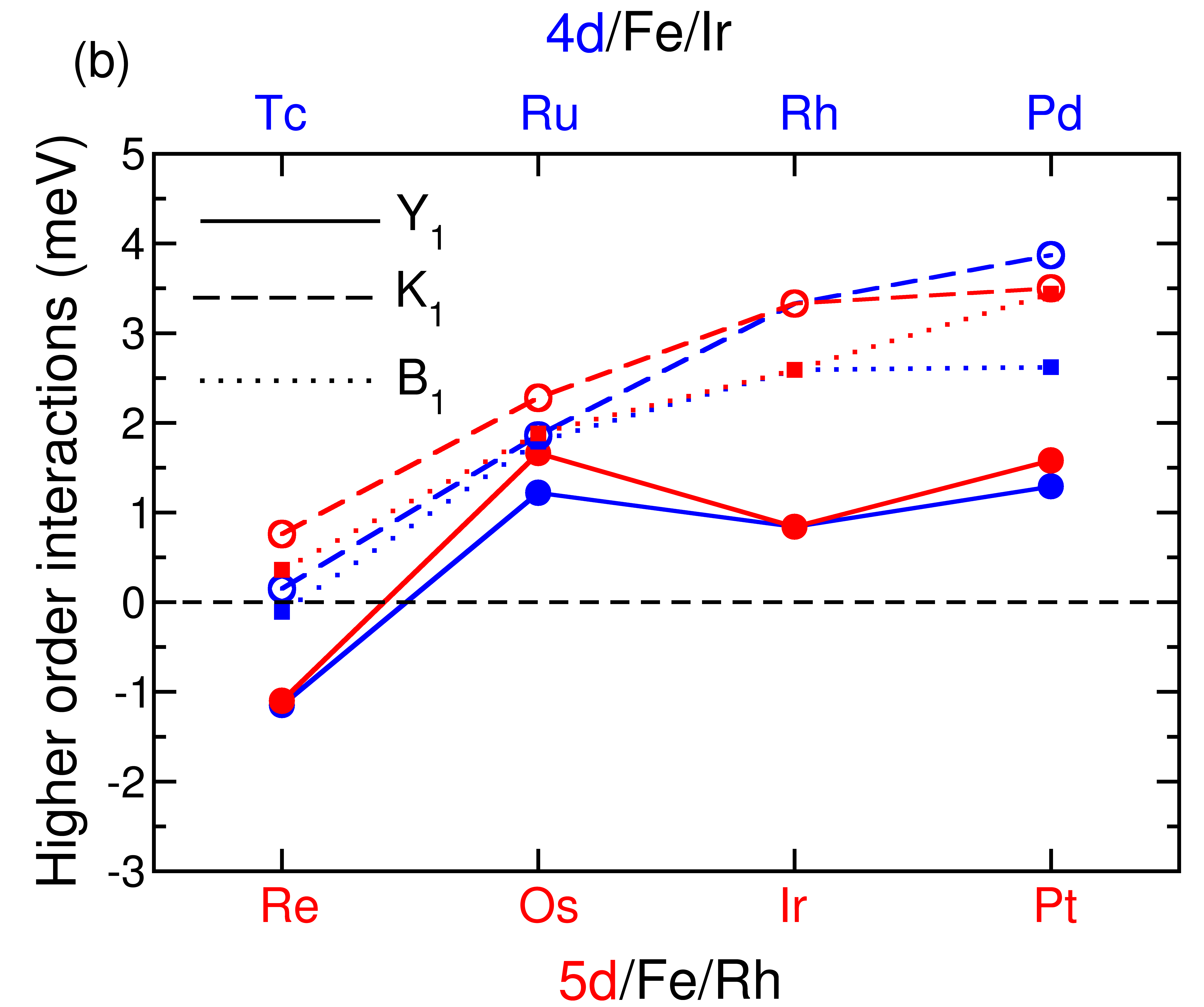}
	\caption{Trends for $4d$/Fe/Ir and $5d$/Fe/Rh trilayers for
	both (a) Heisenberg exchange constants up to the third nearest neighbor as extracted from fitting the corresponding spin spiral dispersions, i.e.~without modification by HOI terms, and (b) HOI parameters $Y_1$ (solid line), $K_1$ (dashed line) and $B_1$ (dotted line) for isoelectronic Fe based trilayers in fcc stacking. 
	The lines connecting the data points serve as a guide to the eye.
	Blue (red) color denotes $4d$/Fe/Ir ($5d$/Fe/Rh) trilayers.}
	\label{J_HOI_4dFeIr_RhFe5d}
\end{figure*}

From the energy dispersions of Fig.~\ref{SS_4dFeIr_RhFe5d} one can evaluate trends for both the Heisenberg and higher-order exchange interactions of these Fe based trilayers. 
First we focus on the pair-wise exchange constants shown in
Fig.~\ref{J_HOI_4dFeIr_RhFe5d}(a). We find that the nearest-neighbor constant $J_1$ rises with increasing number of electrons both in the $4d$ as well as in the $5d$ shell. The trend is very similar to that observed already for
the symmetric $4d$/Fe/$4d$ trilayers, however, the magnitude of $J_1$ is much reduced. This is also a result
of the stronger hybridization since we have used the relaxed interlayer distances for the asymmetric trilayers
(cf.~table \ref{tab:interlayerdistance}).
For the isoelectronic $4d$/Fe/Ir and $5d$/Fe/Rh trilayers the values of $J_1$ are quite similar and
follow the same trend.

\begin{table}[htb]
	\centering
	\caption{Magnetic moments of the asymmetric Fe based trilayers for the ferromagnetic state. All values are given in $\mu_B$.}
	\label{tab:table3}
	\begin{ruledtabular}
		\begin{tabular}{l c c c c }
			$4d$/Fe/Ir& $m_{4d}$ & $m_{\text{Fe}}$ &$m_{\text{Ir}}$  \\ 
			\colrule
			Tc/Fe/Ir & $-0.21$  & $+1.70$   & $+0.08$  \\
			Ru/Fe/Ir & $-0.26$ & $+1.75$  & $-0.03$  \\
			Rh/Fe/Ir & $+0.23$ & $+2.43$ & $-0.00$\\
			Pd/Fe/Ir &$+0.37$ &$+2.81$ & $+0.09$\\
			\colrule
			$5d$/Fe/Rh& $m_{5d}$&$m_{\text{Fe}}$ & $m_{\text{Rh}}$\\ 
			\colrule
			Re/Fe/Rh &$-0.14$ &$+2.11$ &$+0.42$   \\
			Os/Fe/Rh & $-0.24$ & $+1.93$&$+0.20$ \\
			Ir/Fe/Rh & $-0.00$ & $+2.43$ & $+0.23$\\
			Pt/Fe/Rh & $+0.30$&$+2.89$ &$+0.38$\\
			 \end{tabular} 
	\end{ruledtabular}
\end{table}	

The same is true for the next-nearest and third-nearest neighbor constant $J_2$ and $J_3$, respectively. However, their qualitative trend is considerably different as compared to the one found for $J_1$. While $J_1$
mediates a FM coupling between nearest neighbors in all cases, $J_2$ displays an AFM behaviour throughout the whole series and $J_3$ even experiences a change of sign as one goes from Tc to Ru (Rh to Pd) and the respective isoelectronic counterparts. These findings support our conclusion from the energy dispersions, namely that one is dealing with largely exchange frustrated systems. It is also clear from Fig.~\ref{J_HOI_4dFeIr_RhFe5d}(a) that the values for the NN constant $J_1$ span a larger order of magnitude ranging up to about 11~meV 
for Pd/Fe/Ir than for $J_2$ and $J_3$ which only exhibit a variation of 5 to 6 meV.

\begin{figure*}[htbp]
	\centering
	\includegraphics[scale=0.4,clip]{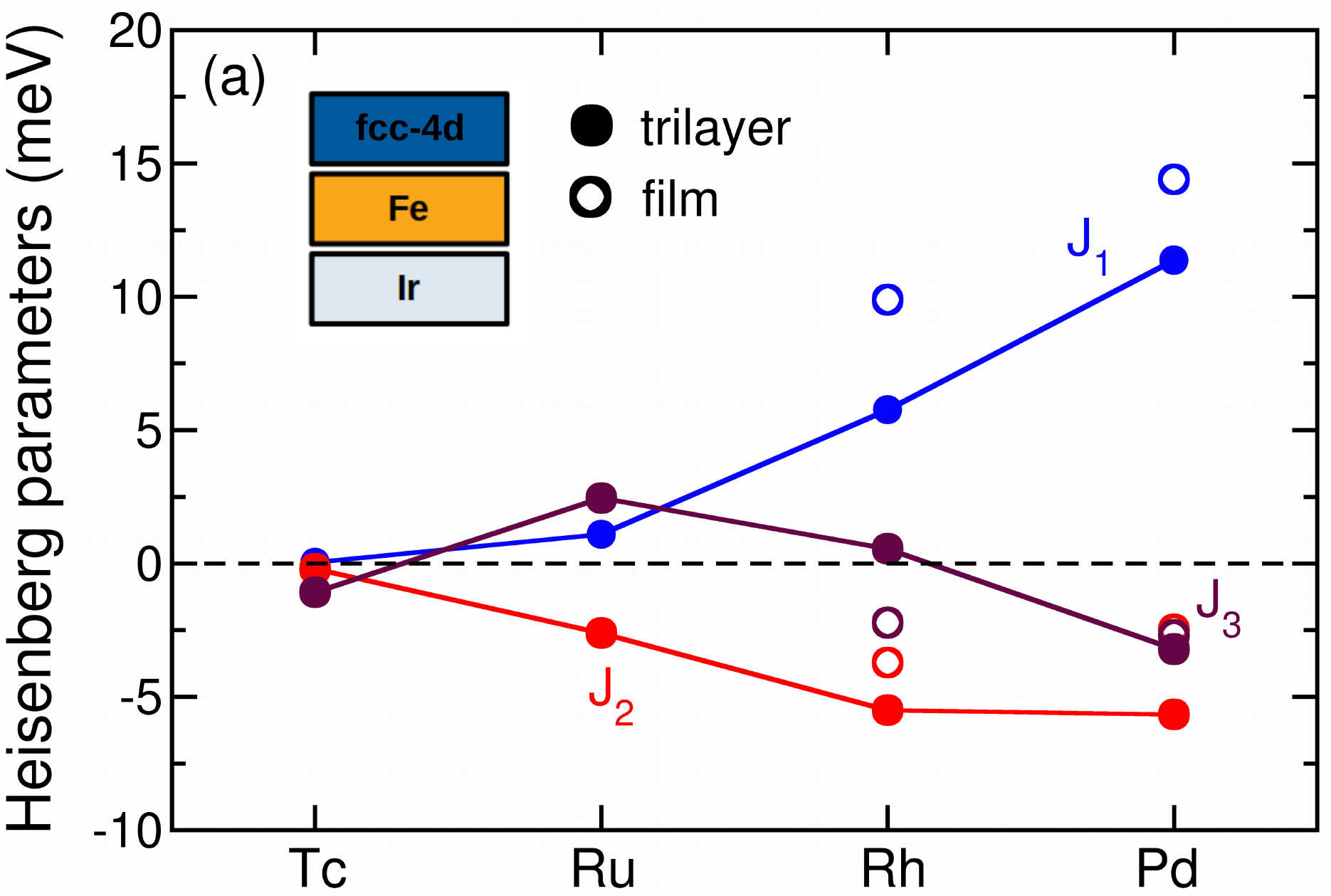}
	\includegraphics[scale=0.4,clip]{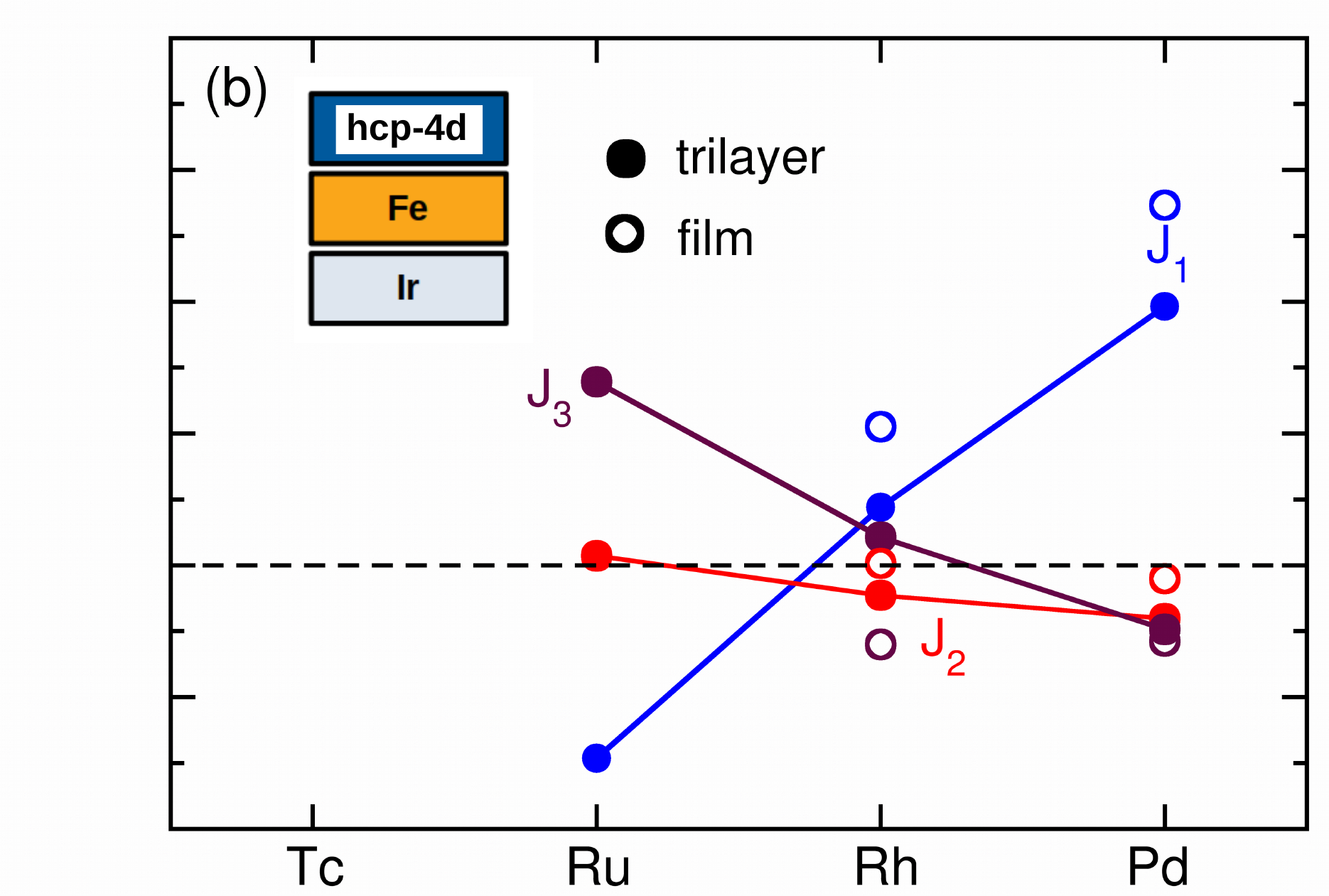}
	\includegraphics[scale=0.4,clip]{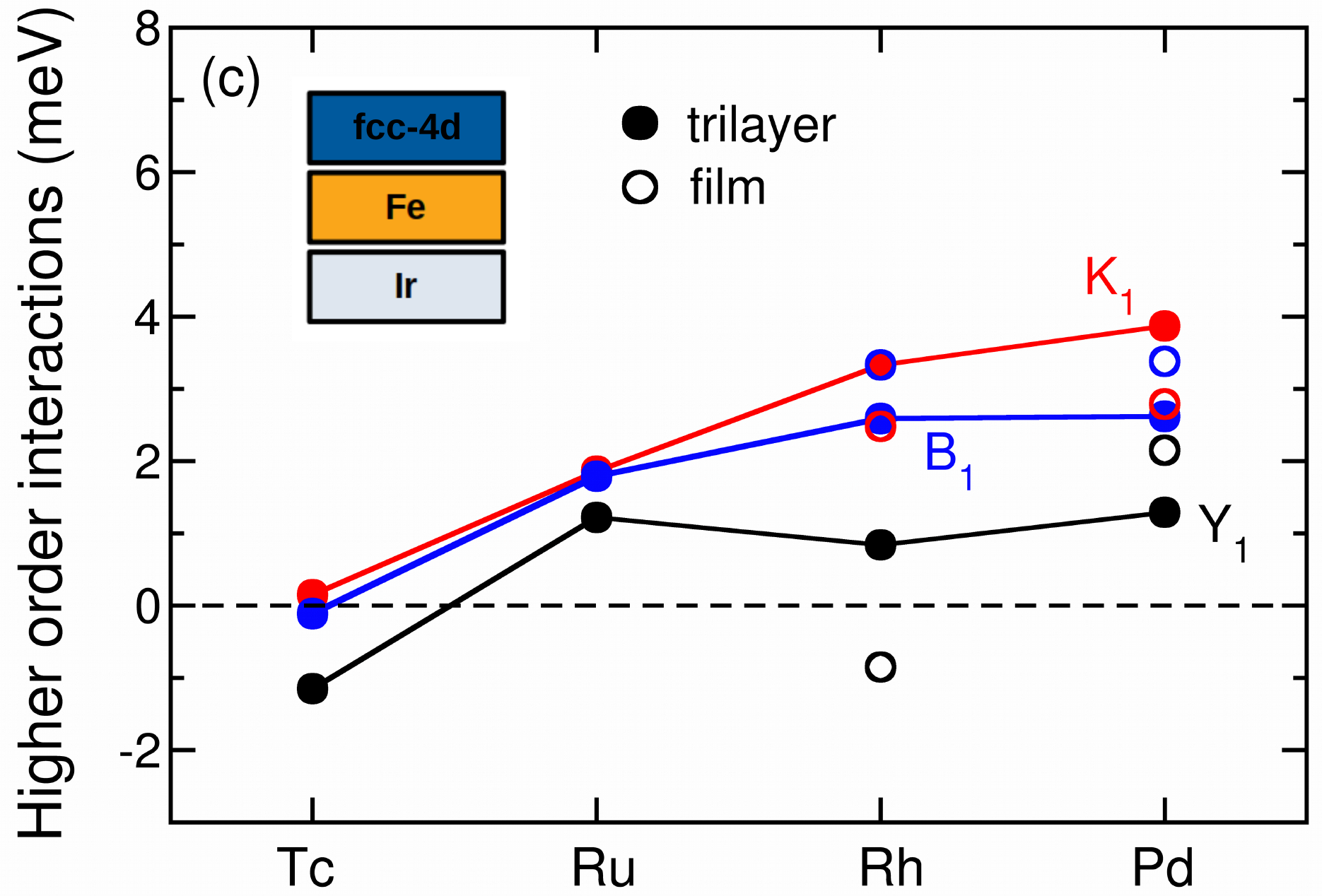}
	\includegraphics[scale=0.4,clip]{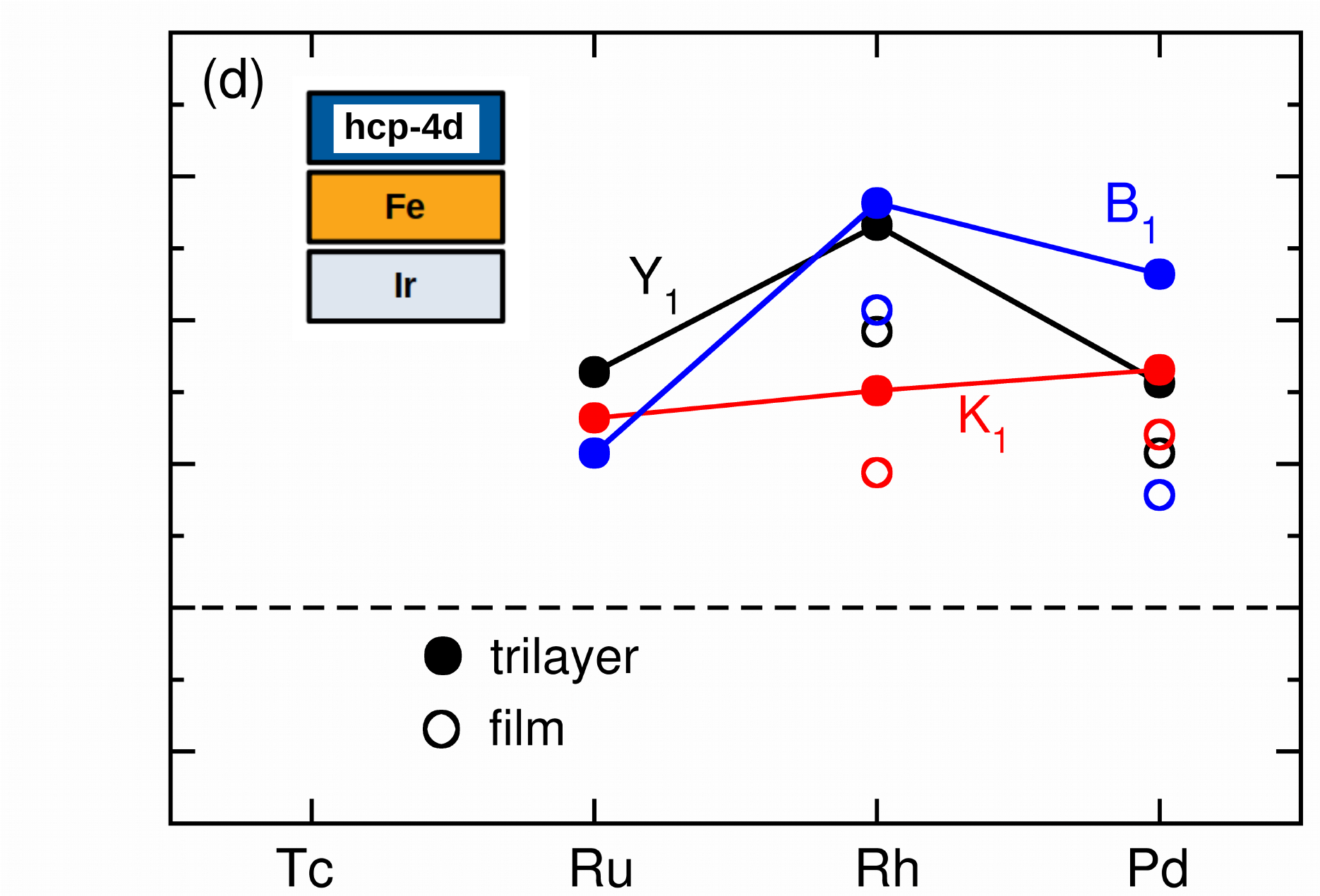}
	\caption{Trend of (a,b) the first three Heisenberg exchange parameters as extracted from fitting the corresponding 
	spin spiral dispersions, i.e.~without modification by HOI terms, and (c,d) the HOI parameters for fcc- and 
	hcp-$4d$/Fe/Ir trilayers (filled circles)
	as well as for the film systems Rh/Fe/Ir(111) and Pd/Fe/Ir(111) (open circles). 
	The lines connecting the data points serve as a guide to the eye.}
	\label{J__HOI_Fe_fcc_hcp_Film}
\end{figure*}

We now turn to the HOI parameters of isoelectronic Fe based trilayers (Fig.~\ref{J_HOI_4dFeIr_RhFe5d}(b))
calculated from the energy differences between 
multi-$Q$ and the respective spin spiral states according to Eqs.~(\ref{eq:B1}-\ref{eq:K1}). One sees a slightly different trend than for the Heisenberg exchange constants. The three-site four spin term, $Y_1$, first rises moving from Tc/Fe/Ir 
(Re/Fe/Rh) to Ru/Fe/Ir (Os/Fe/Rh) thereby experiencing additionally a change of sign before remaining nearly constant at 
$\approx$ 1 meV for the rest of the series. Here, the qualitative behavior is observed to be the same for $4d$/Fe/Ir and 
$5d$/Fe/Rh trilayers as well. In contrast, both the biquadratic and four-site four spin term only exhibit positive values 
(except for Tc/Fe/Ir) which rise with increasing $d$ band filling 
(Fig.~\ref{J_HOI_4dFeIr_RhFe5d}(b)). Additionally, both display higher maximum values of up to about 3.9 meV for Pd/Fe/Ir than for $Y_1$ which peaks at 1.7 meV for Os/Fe/Rh. 

\textbf{Effect of stacking sequence.} Next we take a look at the influence of varying the stacking sequence of the atomic layers upon both the Heisenberg exchange and HOI parameters. This effect is investigated for the example of the $4d$/Fe/Ir trilayer series and the 
corresponding ultrathin film systems Rh/Fe/Ir(111) and Pd/Fe/Ir(111) which have been studied 
experimentally~\cite{Romming2018,Romming2013,Romming2015,Hanneken2015,Leonov_2016}.

Fig.~\ref{J__HOI_Fe_fcc_hcp_Film} shows the trends of all pair-wise and higher-order
exchange parameters for fcc- and hcp-$4d$/Fe/Ir trilayers and the film systems mentioned.
As for the fcc-$4d$/Fe/Ir trilayers (Fig.~\ref{J__HOI_Fe_fcc_hcp_Film}(a)), there is an increase of $J_1$ with band 
filling in the adjacent $4d$ layer (Fig.~\ref{J__HOI_Fe_fcc_hcp_Film}(b)).
However, for the hcp stacked trilayers we find a change of sign of $J_1$ from Rh/Fe/Ir to 
Ru/Fe/Ir~\footnote{Note that we could not converge calculations for the hcp-Tc/Fe/Ir trilayer.} similar to
that observed for Fe monolayers in hcp stacking on Ir(111)~\cite{Hardrat2009}. 
The second and third-nearest neighbor exchange, $J_2$ and $J_3$, show the same qualitative behavior for the two types
of stacking, i.e.~both drop along the series until they mediate an AFM coupling for both fcc- and hcp-Pd/Fe/Ir 
(see Figs.~\ref{J__HOI_Fe_fcc_hcp_Film}(a,b)). This effect is slightly smaller for 
$J_2$ in the hcp trilayers with values ranging between about 0.4 meV for Ru and 
$-2.0$ meV for Pd than in the fcc systems for which this exchange constant spans a larger order of magnitude with about 
5.5 meV. The opposite is true for $J_3$ whose values vary over a range of more than 9 meV for the case of hcp stacking 
as compared to fcc-$4d$/Fe/Ir trilayers with a considerably smaller range of about 5.6 meV. In particular, for
hcp-Ru/Fe/Ir $J_1$ and $J_3$ are of nearly the same absolute value but opposite sign. 

Figs.~\ref{J__HOI_Fe_fcc_hcp_Film}(c,d) reveal a qualitatively different behavior for the HOI parameters of 
hcp and fcc trilayers. 
Especially the biquadratic and the three-site four spin terms, $B_1$ and $Y_1$, exhibit larger values in 
hcp stacking throughout the series. In addition, they also show a striking maximum of about 5.6~meV and 
5.3~meV, respectively, for hcp-Rh/Fe/Ir. In contrast, $K_1$ which amounts to a large value of about 3.0 meV turns 
out to be the smallest HOI term for this trilayer. The large values of $B_1$ and $Y_1$ are linked to the 
$uudd$ state along $\overline{\Gamma {\rm M}}$ being the ground state of hcp-Rh/Fe/Ir. In contrast, a spin
spiral ground state was observed for fcc-Rh/Fe/Ir (cf.~Fig.~\ref{SS_4dFeIr_RhFe5d}). 

\textbf{Trilayers vs.~film systems.} Now we compare the trilayer results with those obtained for film systems. Concerning the pair-wise
exchange constants (Figs.~\ref{J__HOI_Fe_fcc_hcp_Film}(a,b)), we find that the film systems Rh/Fe/Ir(111)
and Pd/Fe/Ir(111) follow the same trend as the corresponding trilayers. This shows that the magnetic
exchange interaction in the Fe layer is dominated by the hybridization at the interfaces with the
$4d$ and $5d$ transition-metal layers. The stacking dependence of the exchange constants is also
captured in the trilayer calculations.

The values of the HOI parameters for hcp-Rh/Fe/Ir(111) are slightly lower than for the trilayer,
however, they are still larger than both the biquadratic 
as well as the three spin term for fcc-Rh/Fe/Ir(111) (Figs.~\ref{J__HOI_Fe_fcc_hcp_Film}(c,d)). 
The obtained values of the HOI constants explain the origin of the stacking-dependent magnetic ground 
state observed by spin-polarized STM~\cite{Romming2018}. In particular, the change of sign and
large increase of the three-site four spin term, $Y_1$, for hcp-Rh stacking drives the change 
from the spin spiral for fcc to the $uudd$ state along $\overline{\Gamma {\rm M}}$
for hcp-Rh stacking (cf.~Eq.~(\ref{eq:Delta_E_GM})).

A similar combination of HOI parameters has been reported for hcp-Fe/Rh(111)~\cite{Kroenlein2018}
in which an $uudd$ state has been found as magnetic ground state. While $K_1$ is vanishingly small for this system, both 
$B_1$ and $Y_1$ are on the same order of magnitude with maximum values of 4 meV~\cite{Kroenlein2018}. The opposite is true for an fcc Fe monolayer on Ir(111) which is known to exhibit a quadratic nanoskyrmion lattice as magnetic ground state induced by the interplay of a strong four-site four spin interaction and the DMI~\cite{Heinze2011}. As calculated 
in~\cite{Kroenlein2018}, fcc-Fe/Ir(111) is indeed characterized by tiny values of $B_1$ and $Y_1$ 
(see also Table~\ref{tab:table4a}), whereas $K_1$ represents the dominant HOI parameter contrary to both hcp-Fe/Rh(111) 
as well as hcp-Rh/Fe/Ir(111).

\begin{table}[h]
	\centering
	\caption{
	Higher-order exchange constants $K_1$, $B_1$ and $Y_1$ for ultrathin film systems. Values for HOI parameters of 
	fcc-Fe/Ir(111) and hcp-Fe/Rh(111) are taken from Ref.~\cite{Kroenlein2018}. All values are given in meV.}
	\label{tab:table4a}
	\begin{ruledtabular}
		\begin{tabular}{l c c c }
			film system& $K_1$ & $B_1$  & $Y_1$  \\ \colrule
			fcc-Rh/Fe/Ir(111) &2.48 &3.33 & $-0.85$\\
			hcp-Rh/Fe/Ir(111) &1.88 &4.14 &3.84 \\
			fcc-Pd/Fe/Ir(111) &2.79 & 3.38&2.15 \\
			hcp-Pd/Fe/Ir(111) &2.41 &1.57 & 2.15\\
			fcc-Rh/Fe/Rh(111) &3.18 &2.79 &1.11 \\
			hcp-Rh/Fe/Rh(111) &2.48 &5.88 &5.12 \\
			\colrule
			fcc-Fe/Ir(111) & $-1.28$&$-0.24$&$-0.24$\\
			hcp-Fe/Rh(111) & 0.10&3.40&4.00\\
			\colrule
			fcc-Rh/Co/Ir(111) &$-0.58$ & 1.79&$-0.96$ \\
			hcp-Rh/Co/Ir(111) &$-1.01$&0.34&$-1.55$\\
			fcc-Pd/Co/Ir(111) &$-1.41$ &1.60 &$-1.39$ \\
		 \end{tabular} 
	\end{ruledtabular}
\end{table}

For Pd/Fe/Ir(111) the observed magnetic ground state is a spin spiral for both stackings~\cite{Romming2013,Hanneken2015}. 
However, in an applied magnetic field a skyrmion lattice becomes favorable and in the field-polarized state
isolated magnetic skyrmions are metastable~\cite{Romming2013,Dupe2014}. Recently, it has been shown that the
four-site four spin interaction $K_1$ plays an important role for the stability of skyrmions~\cite{Paul2020}.
For a positive sign of $K_1$, observed for the Pd/Fe/Ir trilayers and Pd/Fe/Ir(111) 
films (Figs.~\ref{J__HOI_Fe_fcc_hcp_Film}(c,d)), the energy barriers  
stabilizing individual topological spin structures
such as skyrmions or antiskyrmions are enhanced by about 40 to 60 times $K_1$ amounting to a large enhancement
for Pd/Fe/Ir(111)~\cite{Paul2020}.

\subsection{Co based trilayers}
\label{Co_trilayers}
In this section we investigate the effect of substituting the central magnetic layer by Co, i.e.~we study 
fcc-$4d$/Co/Ir trilayers.
A motivation is the recent observation of magnetic skyrmions in zero magnetic field in atomic Rh/Co bilayers on the Ir(111) surface~\cite{Meyer2019}.  
On the other hand, for Pd/Co/Ir(111) only ferromagnetic domains have been
observed in spin-polarized STM experiments \cite{PhysRevB.86.094427}.
We start again by discussing the energy dispersion of flat spin spirals for fcc-$4d$/Co/Ir trilayers before presenting trends for the Heisenberg exchange and HOI parameters of both trilayers and the corresponding film systems for 
which we have calculated exchange constants and HOI parameters as well.

\begin{figure}[htbp]
	\centering
	\includegraphics[scale=0.48,clip]{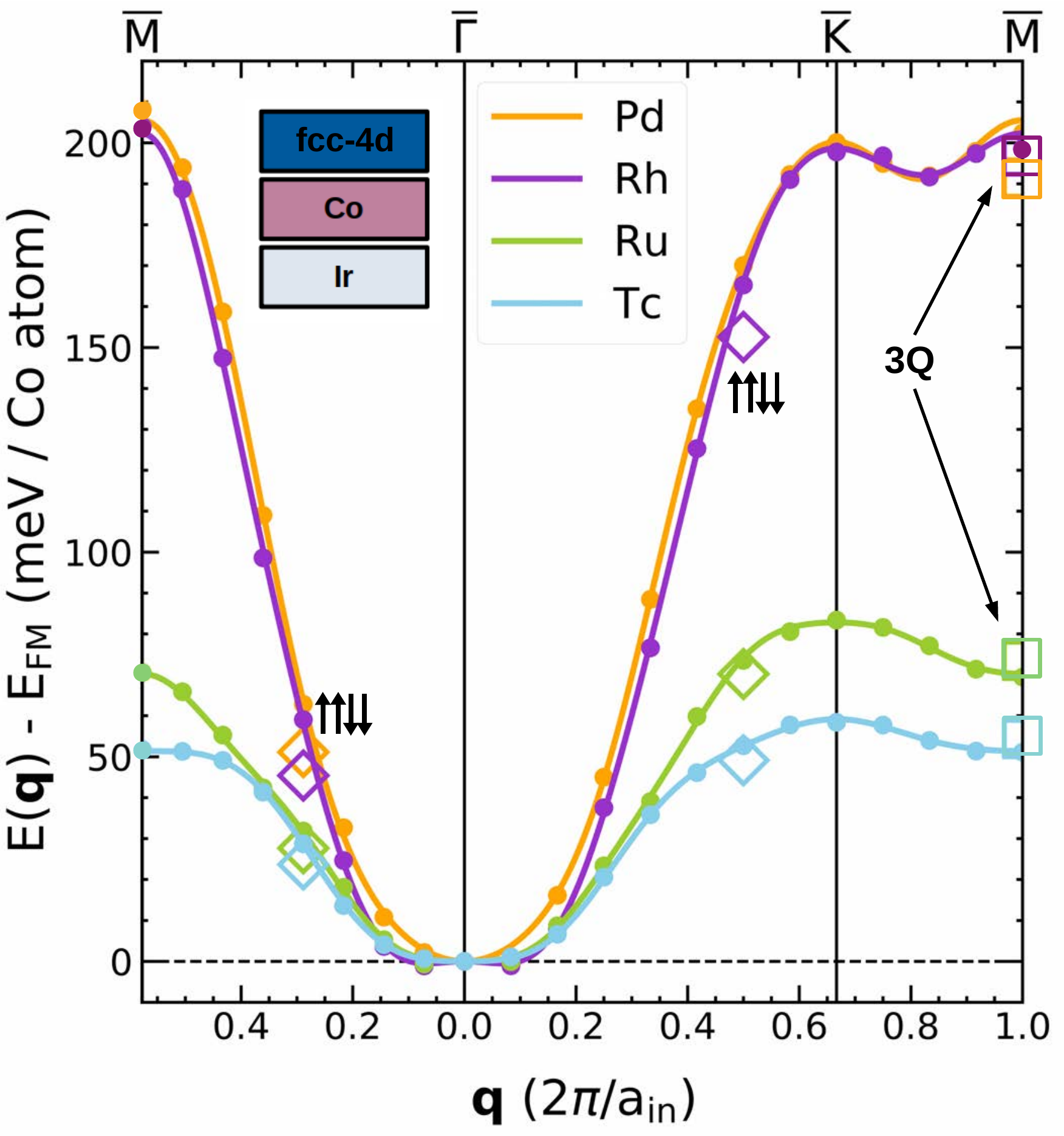}
	\caption{Energy dispersion of flat spin spirals for fcc-$4d$/Co/Ir trilayer systems. Total energies calculated by DFT are marked by filled circles, whereas the solid lines represent fits to the Heisenberg model. Both energies of the 
	\textit{uudd} ($\uparrow \uparrow \downarrow \downarrow$) states as well as the \textit{3Q} state are plotted as empty diamonds and squares at the \textit{q} values of the corresponding single-Q states, respectively.}
	\label{SS_Co}
\end{figure}

The energy dispersion of flat spin spirals calculated along both high symmetry directions of the 2D BZ for Co based trilayers is shown in Fig.~\ref{SS_Co}. In contrast to the previously presented Fe based trilayers, exchange 
frustration effects do not manifest themselves in energy minima of the dispersion
(cf.~Fig.~\ref{SS_4dFeIr_RhFe5d}(a)), but rather in very flat $E(\mathbf{q})$ curves around the $\overline{\Gamma}$-point. 
Furthermore the energy difference between the FM state at the $\overline{\Gamma}$-point and the RW-AFM at the boundary 
of the BZ is quite large for the Co based trilayers compared to the corresponding $4d$/Fe/Ir systems (see 
Fig.~\ref{SS_4dFeIr_RhFe5d}(a)) indicating a stronger FM coupling of the nearest neighbors. This becomes most evident 
for Rh/Co/Ir and Pd/Co/Ir with values of about 200~meV/Co atom (Fig.~\ref{SS_Co}). 
The same behavior of spin spiral energy dispersions has recently been reported for the ultrathin Rh/Co/Ir(111) film 
system based on DFT~\cite{Meyer2019} and the trend of $4d$/Co/Ir(111) is very similar~\cite{MeyerPhD} to that observed 
in the trilayers.  

Fig.~\ref{SS_Co} also shows that the energy differences between the $uudd$ states along both high-symmetry directions 
of the 2D BZ and the corresponding single-Q states are negative in all cases. This observation represents another qualitative difference compared to the Fe based trilayers.  

\begin{table}[htb]
	\centering
	\caption{Magnetic moments of the Co based trilayer systems for the ferromagnetic state. All values are given in $\mu_B$.}
	\label{tab:table5}
	\begin{ruledtabular}
		\begin{tabular}{l c c c c }
			$4d$/Co/Ir& $m_{4d}$ & $m_{\text{Co}}$ &$m_{\text{Ir}}$  \\ 
			\colrule
			Tc/Co/Ir & $-0.12$  & $+1.07$   & $+0.05$  \\
			Ru/Co/Ir & $-0.06$ & $+1.53$  & $+0.16$  \\
			Rh/Co/Ir & $+0.66$ & $+2.05$ & $+0.28$\\
			Pd/Co/Ir &$+0.35$ &$+2.08$ & $+0.43$\\
			\end{tabular}
	\end{ruledtabular}
\end{table}	

Table~\ref{tab:table5} lists both the magnetic moments of the Co atoms as well as the induced moments of the adjacent 
non-magnetic $4d$ and Ir layers in the FM state. As for the $4d$/Fe/Ir trilayers 
(see Table~\ref{tab:table3}), our DFT calculations reveal an antiparallel alignment between the induced magnetic
moments in the $4d$ layers
and the Co atom for the first half of the series and a parallel alignment for the second half. As expected
from Hund's rules, the magnetic moments of the Co atoms turn out to be smaller than the moments 
in the respective Fe based trilayers. Due to increased $3d-4d$ hybridization the Co moment is reduced for 
adjacent layers of $4d$ elements from the beginning of the series. 
For example, Co sandwiched between Tc and Ir exhibits the smallest value of 
1.07 $\mu_B$, while Co in between Pd and Ir has the largest one with 2.08 $\mu_B$. 

\begin{figure*}[htbp]
	\centering
	\includegraphics[scale=0.5,clip]{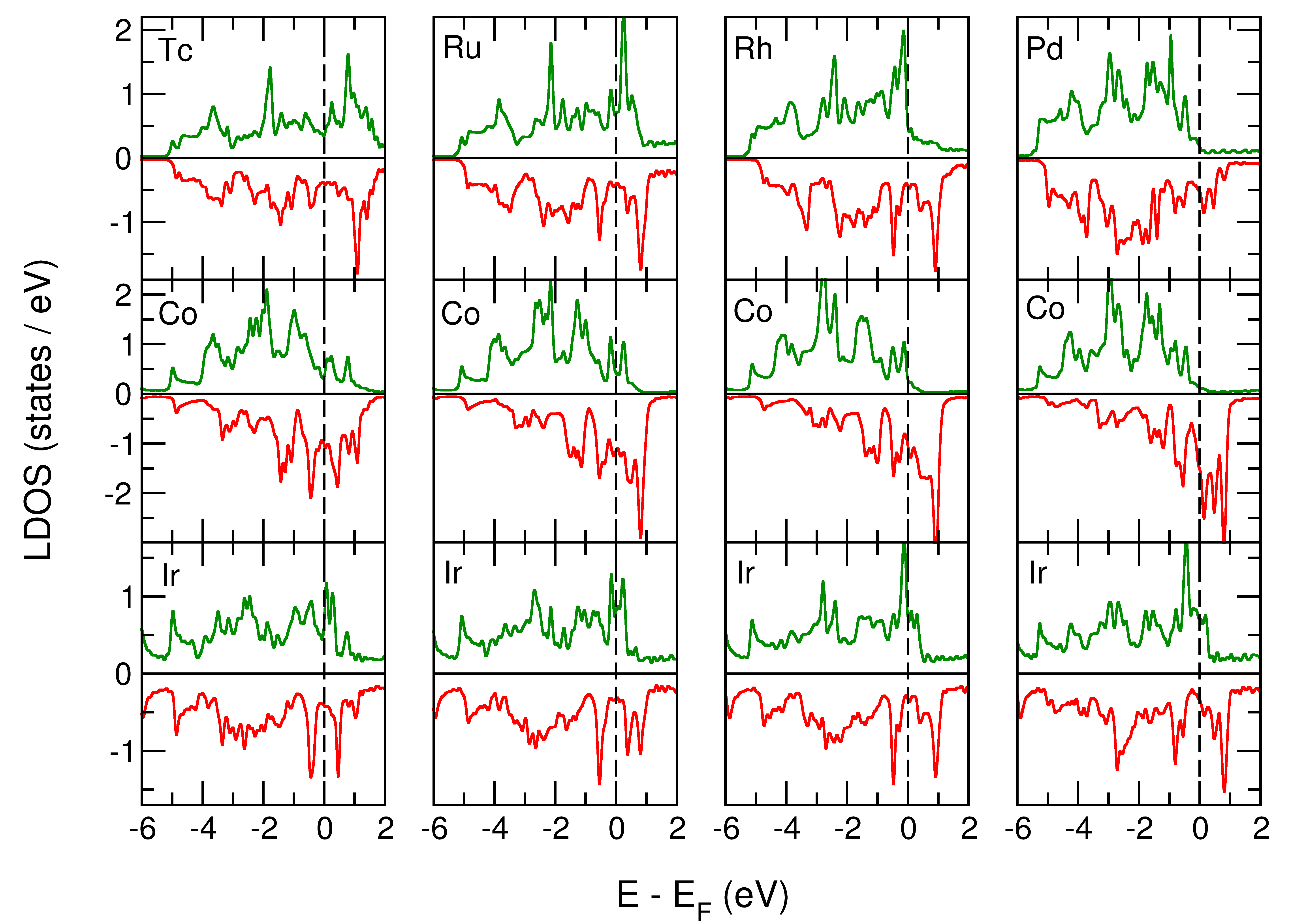}
	\caption{Local density of states (LDOS) of the fcc-$4d$/Co/Ir trilayers. The upper panels show the LDOS of the
	$4d$ overlayer, the middle panels the LDOS of the Co layer, and the lower panels that of the Ir layer. Majority
	and minority spin channels are represented in green and red color, respectively.}
	\label{LDOS_4dCoIr}
\end{figure*}

The spin-resolved local density of states for the $4d$/Co/Ir trilayers (Fig.~\ref{LDOS_4dCoIr}) displays the
band filling trend in the $4d$ overlayer and its effect on the hybridization with the Co layer. For the
Tc and Ru overlayer the hybridization in the majority spin channel even shifts Co states above the Fermi 
energy which leads to the relatively strong reduction of the Co magnetic moment. Several of the peaks in
the vicinity of the Fermi energy appear at the same energetic position in all three layers, indicating that
these are hybrid states between all three atom types. It is also visible how these states shift to lower
energies with increasing band filling of the $4d$ band. 
The strong hybridization
effect is the origin of the drastic change of the spin spiral energy dispersion seen in Fig.~\ref{SS_Co}
for Tc/Co/Ir and Ru/Co/Ir. For Rh and Pd overlayers there is still a significant effect on the Co LDOS,
however, the majority spin band of Co is almost completely filled typical for Co. For Rh/Co/Ir, the majority 
spin LDOS is still enhanced at the Fermi level and the peaks just below the Fermi energy are hybrid states.  

\begin{figure*}[htbp]
	\centering
	\includegraphics[scale=0.4,clip]{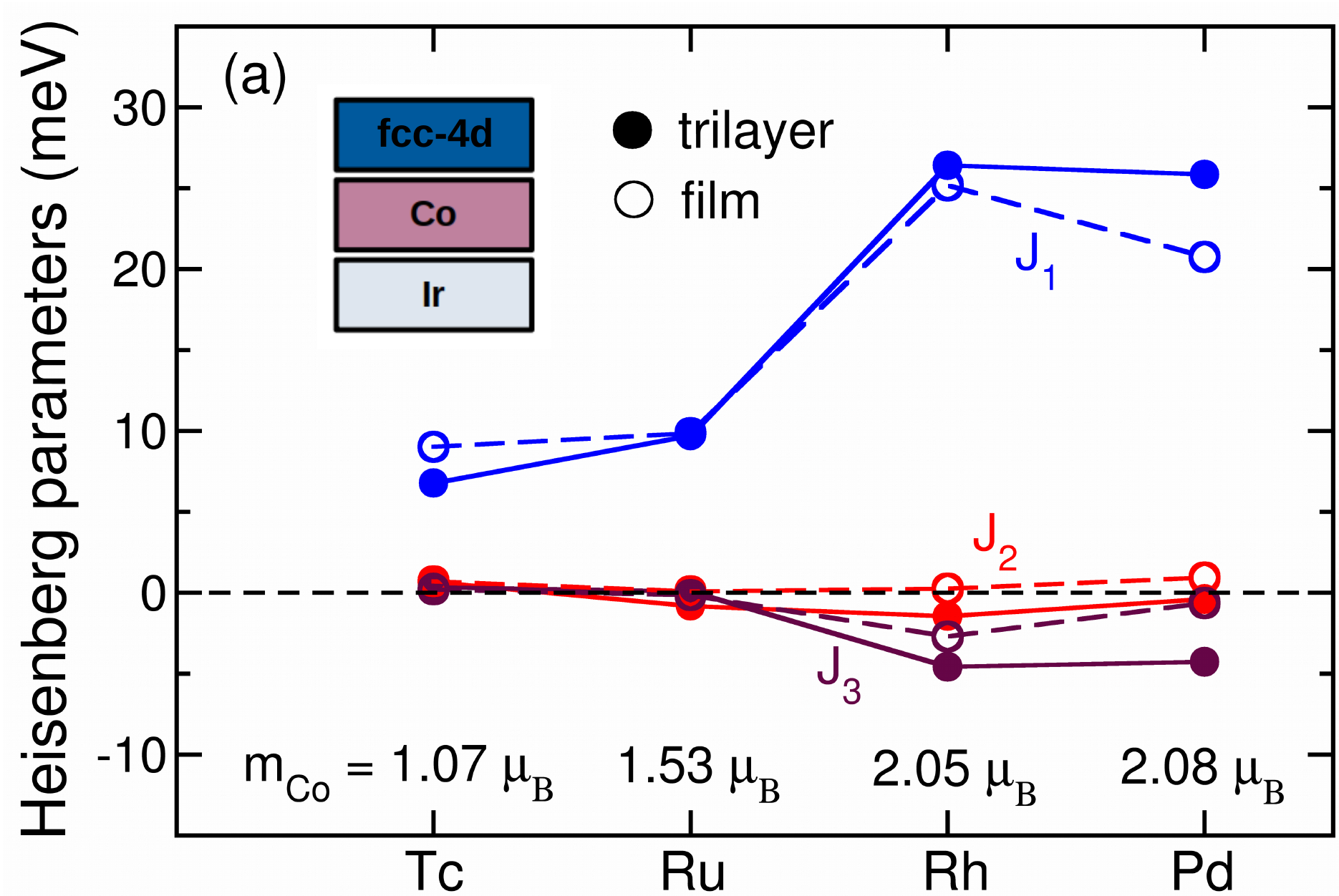}
	\includegraphics[scale=0.4,clip]{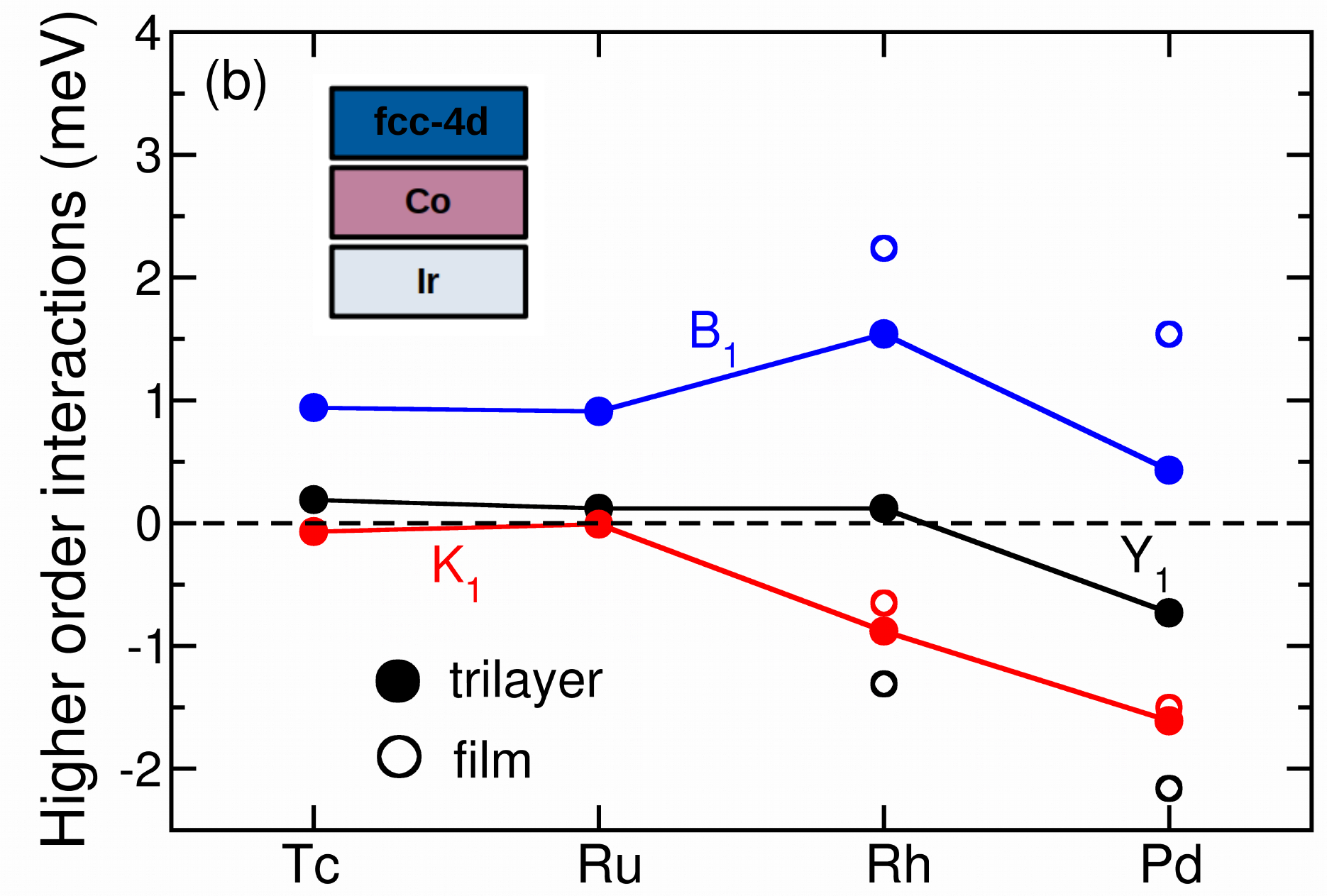}
	\caption{Trends for both (a) Heisenberg exchange constants up to the third nearest neighbor as extracted from fitting the corresponding spin spiral dispersions, i.e. without modification by HOI terms, and (b) the HOI parameters $Y_1$, $K_1$ and $B_1$ for fcc-$4d$/Co/Ir trilayers (filled circles and solid lines) and film systems Rh/Co/Ir(111) and Pd/Co/Ir(111)
	(open circles and dashed lines). The lines connecting the data points serve as a guide to the eye.}
	\label{J_HOI_Co}
\end{figure*}

In accordance with the observations from the spin spiral dispersions, Fig.~\ref{J_HOI_Co}(a) presents large values for the exchange constant $J_1$ of both Rh/Co/Ir and Pd/Co/Ir trilayers as well as the corresponding film systems
Rh/Co/Ir(111) and Pd/Co/Ir(111).
Starting from about 6.8 meV for Tc/Co/Ir, the NN constant peaks at about 26.4 meV for Rh/Co/Ir.
Therefore, the FM exchange coupling between nearest neighbors is much stronger than for the $4d$/Fe/Ir trilayers.
While $J_1$ can be influenced much less in the Co based trilayers by hybridization with the adjacent $4d$ layer
it still allows a reduction by more than a factor of two for Ru and Tc adlayers.  

In contrast to $J_1$, the next-nearest neighbor constant $J_2$ is very small with maximum values of up to about
$-1.5$ meV for the Rh/Co/Ir trilayer and only barely noticeable protruding the zero line for the ultrathin film systems. 
A similar behaviour becomes manifest for $J_3$ with a larger maximum values of about $-4.6$ meV for Rh/Co/Ir and 
$-2.7$ meV for Rh/Co/Ir(111). While the FM pair-wise Heisenberg exchange interaction dominates the 
next-nearest and third nearest neighbor constants mediating mostly an AFM coupling in Co based trilayers, there is
a considerable exchange frustration present which leads to the remarkably flat energy dispersion curve close to the 
FM state especially for Rh/Co/Ir and Rh/Co/Ir(111)~\cite{Meyer2019}.

Comparing further the HOI parameters of Co based trilayers and film systems in 
Fig.~\ref{J_HOI_Co}(b) and Table~\ref{tab:table4a}, one notices a similar trend for $Y_1$, $K_1$ and $B_1$ along the trilayer series at first glance. Up to Ru/Co/Ir the three exchange constants nearly stay constant with $B_1$ showing 
the largest values of 
about 1.5 meV for Rh/Co/Ir and $K_1$ almost vanishing in the case of Tc/Co/Ir and Ru/Co/Ir. $Y_1$ turns out to be 
extremely small as well with maximum values of about 0.2 meV occurring for Tc/Co/Ir. This trend considerably reverses 
for Pd/Co/Ir as $Y_1$ and $K_1$ start dropping below zero and taking significant values of $K_1 \approx -1.6$~meV. 

The two film systems Rh/Co/Ir(111) and Pd/Co/Ir(111) follow a similar trend as the respective trilayers: 
$K_1$ exhibits negative values each with a maximum of 
about $-1.4$ meV for Pd/Co/Ir(111), whereas the biquadratic term is positive and on the same order of magnitude for both systems. Comparing further two different stackings of the Rh top layer in the Rh/Co/Ir(111) system, one finds a negative value of $K_1$ which is nearly twice as large for hcp-Rh as compared to fcc-Rh (see Table~\ref{tab:table4a}). The opposite is true for the biquadratic term whose size is approximately five times larger in the fcc case. The negative value
of $K_1$ for Rh/Co/Ir(111) considerably lowers the stability of skyrmions \cite{Paul2020} which have been observed
in these films at zero magnetic field~\cite{Meyer2019,Perini2019}.  

For comparison, we have studied the symmetric trilayer Rh/Co/Rh which is isoelectronic to Rh/Co/Ir.
As seen in Table~\ref{tab:table4}, this approach only causes minor changes for the magnetic exchange constants
similar to our finding for Fe based trilayers. With the exception of the three spin term $Y_1$ that experiences a change 
of sign, the HOI parameters are on the same order of magnitude as the respective values of Rh/Co/Ir. Our calculations reveal a similar behavior for the spin spiral dispersion of Rh/Co/Rh and Rh/Co/Rh(111) as well \cite{MeyerPhD} (not shown).

\section{Conclusion}
\label{conclusion}

We have presented a systematic first-principles study about the behavior of magnetic exchange interactions in transition-metal trilayers with a central Fe or Co layer and compared the results with those obtained for the
corresponding ultrathin films on surfaces. We included not only the pair-wise Heisenberg exchange interaction, but also higher-order contributions arising in fourth order perturbation theory from the Hubbard model. 
We investigated the effect of the stacking sequence of the trilayers on pair-wise exchange constants and
on the biquadratic interaction and the three-site and four-site four spin interactions. Trends for these magnetic 
exchange constants were obtained by replacing the outer layers of the respective trilayer with different elements from 
the $4d$ and $5d$ series.
Since there is a strong $3d-4d$ and $3d-5d$ hybridization, the band filling has a large impact on both pair-wise
and higher-order exchange constants.
For comparison we analyzed trends for these exchange parameters in symmetric $4d$/Fe/$4d$ systems. 

We find that the exchange interactions in Fe and Co layers can be drastically modified due to the hybridization 
between $3d$, $4d$, and $5d$ bands in these sandwich structures. The trend of the interactions with band filling
does not depend on using isoelectronic $4d$ and $5d$ transition metals. We find a good agreement of the trends 
between trilayer and film calculations highlighting that the hybridization at the two interfaces is the main
origin of the modified exchange interactions.

In Fe based trilayers we find that the
magnetic ground state strongly varies upon replacing the adlayers and the ferromagnetic or row-wise antiferromagnetic
state as well as spin spiral states can occur. Upon varying the stacking sequence from fcc to hcp, we find that
due to higher-order exchange interactions even a superposition state of two spin spirals, the $uudd$ (2Q) state, can
be stabilized in hcp-Rh/Fe/Ir and hcp-Rh/Fe/Rh trilayers. Our calculations of the higher-order exchange constants
reveal that this effect, which has been previously observed for Rh/Fe/Ir(111)~\cite{Romming2018}, is dominated by the
change of the three-site four spin interaction. For the Rh/Fe/Ir trilayers, the higher-order exchange constants can 
even be of larger or comparable value than the pair-wise exchange interactions. The largest values of the higher-order
constants are obtained for the film system Rh/Fe/Rh(111).
For Pd/Fe/Ir trilayers and the corresponding film system Pd/Fe/Ir(111)
we find large positive values of the four-site four spin interaction constant which will lead to an enhanced stability 
of magnetic skyrmions.

In Co based trilayers the hybridization with the adjacent transition-metal layers can also lead to a large variation 
of pair-wise exchange constants, however, the magnetic ground state remains the ferromagnetic state and the 
nearest-neighbor exchange dominates. Nevertheless, a 
strong frustration of exchange interactions can occur. This can lead to magnetic skyrmions being stable in zero magnetic
field as
in Rh/Co/Ir(111)~\cite{Meyer2019}. The values of the higher-order exchange constants are mostly smaller
than in the Fe based trilayers. Importantly, we find that the four-site four spin interaction is negative for all
considered Co based trilayers and film systems which will lead to a decreased stability of skyrmions in these systems.

\section*{Acknowledgments} 

We gratefully acknowledge computing time at the supercomputer of the North-German Supercomputing Alliance (HLRN) and financial support from the Deutsche Forschungsgemeinschaft (DFG, German Research Foundation) via project no.~418425860 (HE3292/13-1).

\appendix
\section{Comparison of methods for HOI}
\begin{table}[htb]
	\centering
			\caption{Comparison of the HOI Parameters $K_1$ and $B_1$ (in meV) for selected trilayer (in fcc-stacking) and film systems
	calculated from a a mixture of {\tt FLEUR} (\textit{uudd} states, spin spirals) and VASP (\textit{3Q} states) denoted by the superscript `mix' and {\tt FLEUR} only. 
	Note that $Y_1$ is evaluated from the energy difference between the two \textit{uudd} states which have only been calculated using {\tt FLEUR}. Therefore, only one 
	value is given. All values are given in meV.}
	\label{tab:table1}
	\begin{ruledtabular}
        \begin{tabular}{l c c c c c}
		& $K_{1}^{\text{mix}}$ &$K_{1}^{\tt{FLEUR}}$ & $B_{1}^{\text{mix}}$ & $B_{1}^{\tt{FLEUR}}$ & $Y_1$ \\ 
			\colrule
			Rh/Fe/Ir & 3.33  & 3.71   & 2.59  & 3.35 & 0.84 \\
			Pd/Fe/Ir & 3.87 & 4.30   & 2.62 & 3.49  &1.29 \\
			\colrule
			fcc-Rh/Fe/Ir(111) & 2.12  & 2.48 & 2.60  & 3.33 &$-0.85$ \\
			hcp-Rh/Fe/Ir(111) &1.94 & 1.88   & 4.27  & 4.14 &3.84 \\
		    \colrule
			Rh/Co/Ir & $-1.15$  & $-0.88$   &  0.98  & 1.54 &0.12 \\
			Pd/Co/Ir & $-1.75$ & $-1.61$ &  0.15 & 0.43 &$-0.73$ \\
			\colrule
			fcc-Rh/Co/Ir(111) & $-0.79$  & $-0.58$ &  1.38  & 1.79 &$-0.96$ \\
			fcc-Pd/Co/Ir(111) & $-1.53$ & $-1.41$ &  1.36 & 1.60 & $-1.39$\\
		\end{tabular} 
	\end{ruledtabular}
\end{table}
In this section we show for selected trilayers and film systems to which extent a mixture of both DFT codes -- the all-electron FLAPW method as implemented in 
{\tt FLEUR} and the PAW method from {\tt VASP} which is based on pseudopotentials -- can 
be used to obtain the trends of the HOI parameters. 
We have chosen two systems from the Fe based and Co based series for which the
corresponding film systems have already been 
studied experimentally~\cite{Romming2018,Meyer2019, Romming2013, Romming2015, Hanneken2015, Leonov_2016,PhysRevB.86.094427}.

We have calculated the energy differences between the two $uudd$ states
and the corresponding spin spiral states, $\Delta E_{\frac{1}{2}\overline{\Gamma\text{M}}}$
and $\Delta E_{\frac{3}{4}\overline{\Gamma\text{K}}}$ 
(cf.~Eqs.~(\ref{eq:Delta_E_GM}-\ref{eq:Delta_E_GK})),
using {\tt FLEUR} while the energy difference between the RW-AFM and the \textit{3Q} state, Eq.~(\ref{eq:Delta_E_3Q}), has either been obtained
with {\tt FLEUR} or alternatively with {\tt VASP}. Using Eqs.~(\ref{eq:B1}-\ref{eq:K1})
we have obtained the biquadratic term, $B_1$, the three-site four spin term, $Y_1$, and the four-site four spin constant, $K_1$, using only the energy 
differences from {\tt FLEUR} (table \ref{tab:table1}). 
We can also use the energy difference 
$\Delta E_{\overline{\text{M}}}$ obtained via {\tt VASP}, which is computationally
less demanding, and $\Delta E_{\frac{1}{2}\overline{\Gamma\text{M}}}$
and $\Delta E_{\frac{3}{4}\overline{\Gamma\text{K}}}$ from {\tt FLEUR}.
As seen in table~\ref{tab:table1} the biquadratic and four-site four spin terms change 
upon using this mixed approach. Note that only one value of $Y_1$ is given in
table \ref{tab:table1} since it is obtained according to Eq.~(\ref{eq:Y1})
from the difference of $\Delta E_{\frac{1}{2}\overline{\Gamma\text{M}}}$ and $\Delta E_{\frac{3}{4}\overline{\Gamma\text{K}}}$.

For the Fe based trilayers, one sees that the values for the four spin term $K_1$ are by about 10\% larger for the all-electron FLAPW method ($K_{1}^{\tt{FLEUR}}$) than for the mixture of both DFT codes ($K_{1}^{\text{mix}}$). However, the sign of $K_1$ is the same 
in both methods for the Rh/Fe/Ir and Pd/Fe/Ir trilayer. The same holds for the biquadratic term $B_1$ albeit the values calculated using {\tt FLEUR} are larger by about 20\% than for 
the mixed method. A similar trend is obtained for the film systems Rh/Fe/Ir(111). 
Therefore, we conclude that the mixed approach provides reliable values of the 
HOI parameters even if the less demanding approach is used in which the noncollinear
\textit{3Q} state is calculated via {\tt VASP}.

For the two Co based trilayers, 
the sign of $K_1$ and $B_1$ is again the same in both approaches. However, the
relative differences are larger and there is no clear trend. 
This becomes not only noticeable for Rh/Co/Ir where the deviation of $B_1$ from {\tt FLEUR} amounts to 36\%, but also for $B_1$ in the Pd/Co/Ir trilayer showing a value which is by 65\% larger for {\tt FLEUR} than for the mixed method (Table~\ref{tab:table1}). Interestingly, 
the deviations are smaller for the Co based ultrathin film systems as compared to the trilayers (see Table~\ref{tab:table1}). While the maximum deviation amounts to 36\% for $K_1$ in case of fcc-Rh/Co/Ir(111), the differences become very small for both $K_1$ as well as $B_1$ in fcc-Pd/Co/Ir(111). Hence we conclude that in principle the combined approach of {\tt FLEUR} and {\tt VASP} yields acceptable results for calculating HOI parameters both in Fe based as well as Co based ultrathin film systems.
However, it is critical to use the same number of Ir substrate layers for the computation 
of energy dispersions of spin spirals and multi-$Q$ states.  

Summarizing the results of our calculations, we conclude that a combined approach of two different DFT codes -- {\tt FLEUR} and {\tt VASP} -- represents an acceptable and
computationally less demanding way of calculating trends of HOI parameters in Fe based trilayers and film systems. Therefore, we used this approach for the other Fe based
trilayers as described in section \ref{Comp_methods}. For the Co based trilayers the
deviations were larger and therefore we used the {\tt FLEUR} code to calculate all 
required total energy differences. While the deviations were reasonable for the Co
based film system we still used only {\tt FLEUR} in order to be consistent with the
trilayer calculations.

\section{Energy differences multi-$Q$ vs.~single-$Q$ states for evaluation of HOI terms}
In the main part of the paper we discussed the trends of HOI constants in Fe and Co 
based trilayers mainly by analysing the data on figures.
For the sake of completeness, we list the energy differences of multi-$Q$ vs.~single-$Q$ states for all studied systems within the paper - trilayers as well as ultrathin film systems (see Table~\ref{tab:tableEnergies}). The respective HOI terms given in the main text were then calculated according to Eqs.~(\ref{eq:B1}-\ref{eq:K1}).  
\begin{table}[htbp]
	\centering
	\caption{ 
	Higher-order exchange constants $K_1$, $B_1$ and $Y_1$ for asymmetric and symmetric Fe and Co based trilayers. 
	The corresponding energy differences of the three multi-Q and the corresponding single-Q states are listed in table \ref{tab:tableEnergies}. 
	All values are given in meV.}
	\label{tab:table4}
	\begin{ruledtabular}
		\begin{tabular}{l c c c }
			$4d$/Fe/Ir & $K_1$ & $B_1$  & $Y_1$  \\ \colrule
			fcc-Tc/Fe/Ir & 0.15  & $-0.11$   & $-1.15$  \\
			fcc-Ru/Fe/Ir  &1.86  & 1.79  & 1.22  \\
			hcp-Ru/Fe/Ir &2.64  & 2.15 & 3.28\\
			fcc-Rh/Fe/Ir &3.33 & 2.59&0.84 \\
			hcp-Rh/Fe/Ir & 3.02&5.63 & 5.32\\
			fcc-Pd/Fe/Ir & 3.87&2.62 & 1.29\\
			hcp-Pd/Fe/Ir &3.31 & 4.64&3.13 \\
			\colrule
			fcc-$5d$/Fe/Rh & &  &  \\
			\colrule
			Re/Fe/Rh & 0.76& 0.36&$-1.10$ \\
			Os/Fe/Rh &2.28 &1.88 & 1.66\\
			Pt/Fe/Rh &3.50 &3.44 & 1.58\\
			\colrule
			$4d$/Fe/$4d$ & & & \\
			\colrule
			fcc-Tc/Fe/Tc  & 0.12&0.35 &$-0.07$ \\
			fcc-Ru/Fe/Ru  &0.81 &0.80 &1.74 \\
			fcc-Rh/Fe/Rh  &3.60 &1.55 &0.90 \\
			hcp-Rh/Fe/Rh  &3.12 &3.81 &4.03 \\
			fcc-Pd/Fe/Pd  &0.94 &1.98 &1.14 \\
			\colrule
			Tc/Co/Ir     &$-0.07$ &0.94 & 0.19 \\
			Ru/Co/Ir     &$-0.01$ &0.91 & 0.12 \\
			fcc-Rh/Co/Ir &$-0.88$ &1.54 & 0.12\\
			Pd/Co/Ir & $-1.61$ & 0.43& $-0.73$\\
			fcc-Rh/Co/Rh  &$-0.66$ &1.26 &$-0.27$ \\
		 \end{tabular} 
	\end{ruledtabular}
\end{table}

\begin{table}[htbp]
	\centering
	\caption{Total energy differences between multi-$Q$ and their corresponding single-$Q$ states defined according to Eqs.~(\ref{eq:Delta_E_3Q}-\ref{eq:Delta_E_GK})
	used to calculate the HOI terms within NN approximation according to
	Eqs.~(\ref{eq:B1}-\ref{eq:Y1})
	in section~\ref{Results_discussion} of the main text (Table~\ref{tab:table4} and~\ref{tab:table4a}). All values are given in meV/Fe atom.}
	\label{tab:tableEnergies}
	\begin{ruledtabular}
		\begin{tabular}{l c c c }
			$4d$/Fe/Ir& $\Delta E_{\frac{1}{2}\overline{\Gamma\text{M}}}$&$\Delta E_{\frac{3}{4}\overline{\Gamma\text{K}}}$ &$\Delta E_{\overline{\text{M}}}$ \\ \colrule
			fcc-Tc/Fe/Ir &6.21 &$-2.96$ &7.16  \\
			fcc-Ru/Fe/Ir  &2.84 &12.60 &22.83   \\
			hcp-Ru/Fe/Ir &$-0.58$ &25.68& 22.18 \\
			fcc-Rh/Fe/Ir &12.95 &19.66&44.88  \\
			hcp-Rh/Fe/Ir &$-19.67$ &22.85 &33.86 \\
		    fcc-Pd/Fe/Ir  &15.29 &25.58 &48.36\\
		    hcp-Pd/Fe/Ir &$-4.59$ &20.45 &43.37 \\
			fcc-Rh/Fe/Ir(111) &9.92 &3.11 &48.72 \\
			hcp-Rh/Fe/Ir(111) &$-16.90$ &13.82 &21.61 \\
			fcc-Pd/Fe/Ir(111) &0.23 &17.41 &36.36 \\
			hcp-Pd/Fe/Ir(111) &4.46 &21.62&22.66\\
			\colrule
			fcc-$5d$/Fe/Rh & & &  \\
			\colrule
			Re/Fe/Rh &9.05 &0.27 &15.90 \\
			Os/Fe/Rh  &4.11 &17.37 &25.50\\
			Pt/Fe/Rh & 7.89&20.51 &47.22  \\
			\colrule
			$4d$/Fe/$4d$ & & &  \\
			\colrule
			fcc-Tc/Fe/Tc  &$-0.19$ &$-0.75$ &3.47 \\
			fcc-Ru/Fe/Ru &$-3.70$&10.19 &3.62 \\
			fcc-Rh/Fe/Rh &18.96 &26.19&41.87 \\
			hcp-Rh/Fe/Rh &$-6.42$&25.82 &32.17 \\
		    fcc-Pd/Fe/Pd &$-4.95$ &4.14 &14.53\\
			fcc-Rh/Fe/Rh(111) &9.87 &18.71 &42.96 \\
			hcp-Rh/Fe/Rh(111) &$-24.17$ &16.77 &30.49 \\
			\colrule
			$4d$/Co/Ir & & & \\
			\colrule
			fcc-Tc/Co/Ir&$-5.07$ &$-3.55$ &3.31 \\
			fcc-Ru/Co/Ir&$4.20$ &$-3.28$ &4.07 \\
			fcc-Rh/Co/Ir&$-13.65$ &$-12.68$ &$-1.82$\\
			fcc-Pd/Co/Ir&$-11.71$ &$-17.53$ &$-10.99$\\
			fcc-Rh/Co/Rh &$-9.27$ &$-11.42$ &1.08\\
			fcc-Rh/Co/Ir(111)&$-8.01$ &$-15.65$ &8.43 \\
			hcp-Rh/Co/Ir(111)&$-3.29$&$-15.65$&$-0.71$\\
			fcc-Pd/Co/Ir(111) &$-12.08$&$-23.22$ &0.95 \\
		 \end{tabular} 
	\end{ruledtabular}
\end{table}


%

\end{document}